\documentclass[twocolumn,prd,floatfix,preprintnumbers,a4paper,nofootinbib,superscriptaddress,groupedaddress]{revtex4-1}
\usepackage[utf8]{inputenc}
\usepackage{amsmath,amssymb}
\usepackage{amsfonts}
\usepackage{bm}
\usepackage{graphics}
\usepackage{float}
\usepackage{amsmath}
\usepackage{amssymb}
\usepackage[normalem]{ulem} 
\usepackage[mathscr]{euscript}
\usepackage{acronym}
\usepackage{float}
\usepackage{enumitem}   

\usepackage[caption=false]{subfig}
\usepackage{lipsum}
\usepackage{mathrsfs}
\usepackage{mathtools}
\usepackage{multirow}
\usepackage{graphicx}
\usepackage{mathtools}
\usepackage{multirow}
\usepackage{graphicx}
\usepackage[usenames,dvipsnames,table,xcdraw]{xcolor}
\usepackage[colorlinks, pdfborder={0 0 0}]{hyperref} 
\usepackage[nameinlink,capitalise]{cleveref}
\usepackage{xspace}
\usepackage{xfp}

\graphicspath{{fig}}

\newcommand{\comment}[1]{{\textcolor{black}{#1}}}

\graphicspath{{fig}}

\newcommand{\macro}[1]{#1}   

\newcommand{\Nrcasesalln}{\macro{142~}}

\newcommand{\Nrcasesall}{\macro{\Nrcasesalln\xspace}}

\newcommand{\chiphen}{\chi_{\rm pheno}}

\begin{document}

\title{A novel ringdown amplitude-phase consistency test}

\author{Xisco Jim\'enez Forteza$^{1,2}$,
Swetha Bhagwat$^{3,4}$,
Sumit Kumar$^{1,2}$,
Paolo Pani$^{3}$}

\affiliation{$^1$ Max Planck Institute for Gravitational Physics (Albert Einstein Institute), Callinstra{\ss}e 38, 30167 Hannover, Germany}
\affiliation{$^2$  Leibniz Universit\"at Hannover, 30167 Hannover, Germany}
\affiliation{$^3$ Dipartimento di Fisica, ``Sapienza" Università di Roma \& Sezione INFN Roma1, Piazzale Aldo Moro 5, 00185, Roma, Italy}
\affiliation{$^4$ Institute for Gravitational Wave Astronomy $\&$ School of Physics and Astronomy, University of Birmingham,
Edgbaston, Birmingham B15 2TT, UK}
 
\begin{abstract}
{The ringdown signal emitted during a binary black hole coalescence can be modeled as a linear superposition of the characteristic damped modes of the remnant black hole that get excited 
during the merger phase. While checking the consistency of the measured frequencies and damping times against the Kerr BH spectrum predicted by General Relativity~(GR) is a cornerstone of strong-field tests of gravity, the consistency of measured excitation amplitudes and phases have been largely left unexplored. 
For a nonprecessing, quasi-circular binary black hole merger, we find that GR predicts a narrow region in the space of mode amplitude ratio and phase difference, independently of the spin of the binary components.
Using this unexpected result, we develop a new null test of strong-field gravity which demands that the measured amplitudes and phases of different ringdown modes should lie within this narrow region predicted by GR. We call this the \emph{amplitude-phase consistency test} and introduce a procedure for performing it using information from the ringdown signal. Lastly, we apply this test to the GW190521 event, using the multimodal ringdown parameters inferred by Capano et al.~(2021)~\cite{Capano:2021etf}. While ringdown measurements errors for this event are large, we show that GW190521 is consistent with the amplitude-phase consistency test. 
Our test is particularly well suited for accommodating multiple loud ringdown detections as those expected in the near future, and can be used complementarily to standard black-hole spectroscopy as a proxy for modified gravity, compact objects other than black holes, binary precession {and eccentricity}.
}
\end{abstract}
\maketitle

\noindent{{\bf{\em Introduction.}}}
A binary black hole~(BBH) ringdown is the gravitational-wave~(GW) signal emitted as the remnant black hole~(BH) formed during a BBH coalescence relaxes towards a stationary configuration~\cite{chandrasekhar:1975zza,teukolsky1,teukolsky2,teukolsky3}. The Kerr metric~\cite{bekenstein:1973ur,carter:1971zc,hawking:1971tu,hawking:1972hy} in Einstein’s general theory of relativity~(GR) uniquely describes this final state. \comment{The ringdown phase is modelled as the evolution of perturbations (set up during the pre-merger stage) on the Kerr metric of the remnant BH. The GW signal emitted is well-approximated as a linear superposition of countably infinite quasi-normal modes~ (QNMs), i.e., exponentially damped sinusoid modes with discrete characteristic complex frequencies, which are the eigenvalues of the radial and angular Teukolsky’s equations~\cite{berti:2009kk,chandrasekhar:1985kt,ferrari:1984zz,kokkotas:1999bd}. Each mode is characterized by its frequency $f_{lmn}$, damping time $\tau_{lmn}$, excitation amplitude $A_{lmn}$, and phase $\phi_{lmn}$,
where the integers ($l,m$) identify the angular dependence of the mode, whereas $n=0,1,2,..$ is the overtone number (see Eq.~\eqref{eq1} for details).}

\comment{While the frequencies and damping times are solely determined by the remnant’s mass and spin, the perturbation condition setup prior to ringdown phase regulate the mode excitation, namely the amplitudes and phases. For a BBH coalescence, there is an intrinsic relation between the initial binary’s parameters and the perturbation condition setup for ringdown, which determine the magnitude of $A_{lmn}$ and $\phi_{lmn}$. 
QNM amplitudes (i.e. excitation factors) account for spacetime's geometry during merger, providing the initial data for ringdown perturbations. 
GW ringdown models implicitly incorporate this into $A_{lmn}$ and $\phi_{lmn}$.}
We can estimate $A_{lmn}$ and $\phi_{lmn}$ using numerical relativity~(NR) simulations~\cite{jimenez-forteza:2016oae,lousto:2016nlp,hofmann:2016yih} corresponding to a set of BBH masses and spins.
Thus, ringdown allows us to check two key predictions of GR in the strong-field regime — a)~the consistency of measured QNM spectrum in the ringdown to the expected Kerr spectrum and, b)~the compatibility of measured mode excitation factors with the prediction obtained from NR BBH coalescences in GR, i.e., consistency with (pre-)merger nonlinear dynamics.

While the former is the focus of a traditional BH spectroscopy, here we concentrate on the latter prospect and devise a novel test of GR called the \emph{ringdown amplitude-phase consistency~(APC) test}. The APC test is based on the observation that, after a suitable normalization, only a narrow region in the mode amplitude-phase space is allowed for a BBH ringdown in GR.
However, the BBH ringdown amplitudes and phases in modified GR ~\cite{Okounkova:2020rqw,Okounkova:2019zjf,East:2021bqk,Elley:2022ept,Lim:2022veo} need not be constrained to lie in this region. We also expect a similar situation when the components of binary system are not Kerr BHs~\cite{Cardoso:2019rvt} (e.g. in neutron-star or more exotic boson-star~\cite{Palenzuela:2017kcg,Helfer:2018vtq,Bezares:2018qwa,Bezares:2022obu} coalescences) within GR. Note that in both these scenarios, the remnant \emph{can} still be a Kerr BH; so while it could pass the standard BH spectroscopy tests, the QNM amplitudes and phases can provide a way to distinguish the event from a GR BBH coalescence based on the nature of the merger.

\comment{A key virtue of this test is that it does not require information from the inspiral phase other than the binary extrinsic parameters.}
It is therefore particularly well suited for massive BBHs, where the inspiral is short and the parameter estimation of the binary intrinsic parameters (e.g. the mass ratio and spins) is uncertain, jeopardizing the accuracy of inspiral-merger-ringdown~(IMR) consistency tests~\cite{LIGOScientific:2021sio}. GW190521 is one such event~\cite{LIGOScientific:2020iuh}. It is also the only event that has been reported to show presence of measurable subdominant angular-mode parameters~\cite{Capano:2021etf}.
\comment{
Furthermore, the parameter estimation for this event yields a primary component mass lying in the pair-instability supernova BH mass gap. This could be an indication for exotic alternatives~\cite{Bustillo:2020syj,Fishbach:2020qag,Bustillo:2021tga} and the data do not exclude some of these possibilities. Thus, while our main scope is to devise the APC test in general, we find GW190521 a particularly interesting test-bed because if such exotic scenarios are possible, they can affect the QNM amplitudes and phases and manifest as a violation of this null test.
}

\noindent{{\bf{ \em Amplitude-phase space of a BBH ringdown.}}}
Our aim here is to show that the mode amplitude-phase space corresponding to a BBH ringdown in GR is constranined to a narrow region. For this, we must first extract the amplitudes and phases of several modes by fitting the NR simulation as a function of BBH masses and spins.  While we mostly focus on non-precessing, quasi-circular BBHs for simplicity, later on we shall also discuss the effect of spin misalignment and eccentricity.

We use the dominant ($l=m=2$, $n=0$) mode as a baseline and work with the intrinsic (independent of sky position, distance, polarization $\psi$) amplitude ratio and phase difference defined as
\begin{align}
    A^{R}_{lmn} \equiv \frac{A_{lmn}}{A_{220}}\,,\qquad
    \delta \phi_{lmn} \equiv \frac{m}{2} \phi_{220}-\phi_{lmn}\,, \label{definitions}
\end{align}
respectively~\cite{CalderonBustillo:2015lrg}. $\delta\phi_{lmn}$ is defined such that it removes the degeneracy between $\phi_{lmn}$ and the coalescence phase $\varphi$ (see Supplemental Material and Ref.~\cite{Baibhav:2020tma}).
The ringdown waveform can be analytically written down as 
\begin{eqnarray} \label{eq1}
h_+ + i\, h_{\times} &=& A_{220} \sum_{lmn}\Big( e^{- i \frac{m}{2}\phi_{220}}  A^{R}_{lmn}e^{i \delta \phi_{lmn}} S_{lmn}(\iota,\varphi) \nonumber\\
& \times& e^{i 2 \pi f_{lmn}t} e^{- t/ \tau_{lmn}}\Big)\,,
\end{eqnarray}
where $S_{lmn}(\iota,\varphi)$ are the spin 2-weighted spheroidal harmonics. For simplicity, we approximate them to \comment{spherical harmonic functions~\cite{berti:2014fga} and as discussed later, this introduces a systematic error} no larger than $1\%$ for spinning remnants with $a_f\lesssim 0.9$ and for the modes considered in this work.
For quasi-circular BBH mergers, $A^{R}_{lmn}$ and $\delta \phi_{lmn}$ are functions of the binary mass ratio $q=m_1/m_2 \geq 1$ and spins. We fit for $A^{R}_{lmn}$ and $\delta \phi_{lmn}$ mode-wise for a set of $\Nrcasesalln$nonprecessing and noneccentric NR simulations from the SXS catalog~\cite{sxscatalog} (other catalogs~\cite{ritcatalog,gatechcatalog} are considered in the Supplemental Material). Our simulation set \comment{is the same used to} calibrate the \texttt{SEOBNRv4HM} model~\cite{cotesta:2018fcv} and spans $q\in\left[1,10\right]$ \comment{and the $z$-component of the spins} $\chi_{1,2}\in\left[-0.9,0.9\right]$. We fit for the $(l,m,n) \in \{(3,3,0), (2,1,0), (4,4,0) \}$ modes and provide ready-to-use analytical fits as a function of $q$ and a post-Newtonian motivated effective spin $\chi_{\rm pheno}(q,\chi_1,\chi_2)$ whose explicit form depends on the mode under consideration~\cite{supplemental:apc}.
%
In the nonspinning limit, these fits are consistent with those obtained in~\cite{borhanian:2019kxt,london:2018gaq,ota:2019bzl,cotesta:2018fcv}, while they correct the results given in~\cite{forteza2020} for the phase difference. 
We refer the reader to Supplemental Material for further details.

\begin{figure}
\includegraphics[width=0.98\columnwidth]{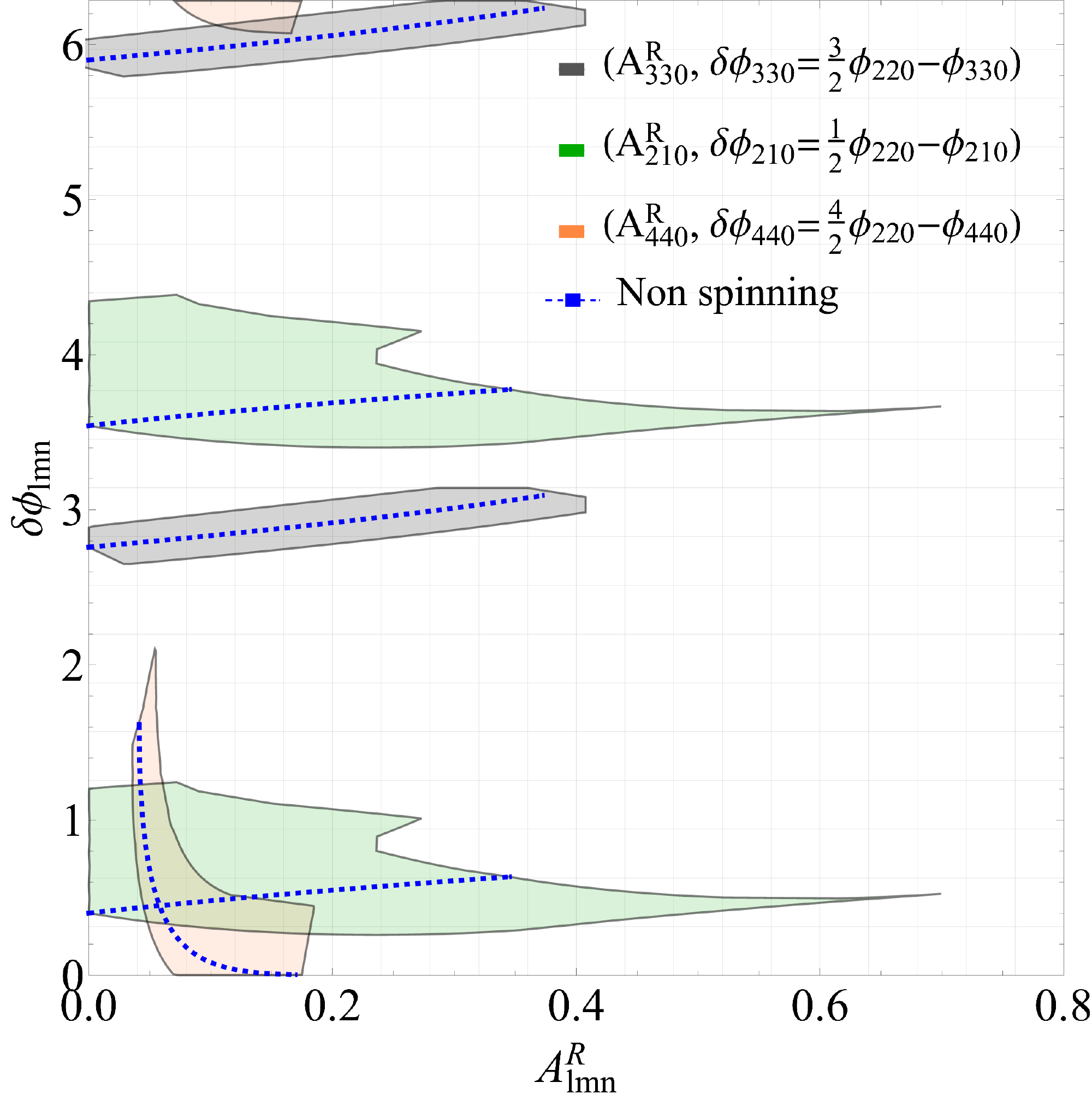}
    \caption{The regions of the $(A^{R}_{lmn},\delta \phi_{lmn})$ plane allowed for BBHs in GR for various modes. They are estimated by fitting the parameters $(A^{R}_{lmn},\delta \phi_{lmn})$ in Eq.~\eqref{eq1} to $\Nrcasesall$nonprecessing waveforms, covering the parameter space $q\in\left[1,10\right]$ and $\chi_{1,2}\in\left[-0.9,0.9\right]$. Given the parity of the odd/even $m$ modes, we plot $\text{mod}(\delta \phi_{lmn},\pi)$ for the ($2,1,0$) and ($3,3,0$) modes and $\text{mod}(\delta \phi_{lmn},2\pi)$ for the ($4,4,0$) mode. The dashed \comment{blue} curves within each region correspond to the nonspinning limit.}
    \label{fig:aratiophasemodes}
\end{figure}

The shaded bands in Fig.~\ref{fig:aratiophasemodes} marks off the region in the $(A^{R}_{lmn},\delta\phi_{lmn})$ space allowed by GR BBH ringdowns as obtained by fitting the amplitudes and phases to our NR dataset using Eq.~\eqref{eq1}.
The dashed curves correspond to $\chi_{1,2}=0$, wherein $A^{R}_{lmn}$ and $\delta\phi_{lmn}$ parametrically depend only on $q$. The shaded region around the dashed curve quantifies the effects of non-zero $\chi_{1,2}$. Given the parity and polarization conventions of the odd/even $m$ modes used in the SXS waveforms, we plot mod$(\delta \phi_{lmn},\pi)$ for the ($2,1,0$) and ($3,3,0$) modes and mod$(\delta \phi_{lmn}, 2 \pi)$ for the ($4,4,0$) mode (see Supplemental Material
).
For the $(3,3,0)$ mode, we find that the GR admissible region in the $(A^{R}_{lmn},\delta\phi_{lmn})$ space is remarkably narrow. This happens because the effects of the progenitor spins are small for $(3,3,0)$ mode. Indeed, $\delta\phi_{330} \in \left[2.68,\pi \right]$ and $A^R_{330} \in \left[0,0.42\right]$ for our entire dataset. \comment{The latter range of values observed for the amplitude ratio $A^R_{330}$ is about a factor two larger than what observed in the inspiral regime~\cite{borhanian:2019kxt}. This shows that higher harmonics are more excited during the highly-dynamical merger and  ringdown regimes.}
However, the effect of non-zero $\chi_{1,2}$ on $A^{R}_{lmn}$ and $\delta \phi_{lmn}$ is substantial for the ($4,4,0$) mode and $(2,1,0)$ mode; this leads to a less-constricted GR-permissible region for these modes. \comment{The small bump of the (2,1,0) shaded area originates from restricting the amplitude $A^R_{210}$ to be positive at small mass ratio and positive spin (see~\cite{Garcia-Quiros:2020qpx} and the Supplemental Material).}

\noindent{{\bf{\em The ringdown APC test.}}}
\emph{All} nonprecessing, quasi-circular BBH ringdown governed by GR must have $A^{R}_{lmn}$ and $\delta \phi_{lmn}$ within in the GR permissible region on the $(A^{R}_{lmn}, \delta \phi_{lmn})$ space i.e., measurements of $A^{R}_{lmn}$ and $\delta \phi_{lmn}$ must lie within the shaded bands in Fig.~\ref{fig:aratiophasemodes}.
We use this to devise our APC null test of strong gravity wherein we check if the measured mode amplitudes and phases in a ringdown signal lies in the narrow GR permissible region. Practically, one can check whether the posterior distributions of the estimated $A^{R}_{lmn}$ and $\delta \phi_{lmn}$ have significant support in the allowed region. Since all quasi-circular BBH mergers must satisfy this constraint, the test is naturally extendable to incorporate a population of observations.

Note that while implementing this test, it is crucial to account for the uncertainty in the ringdown start time with respect to the global peak time $t^{\rm p}$, i.e the time at which the strain amplitude $|h(t)= \Sigma_{lmn} h_{lmn}|$ maximizes~\cite{Bhagwat:2019dtm,forteza2020}. Estimating the time $t^{\rm p}_{lmn}$ at which each $|h_{lmn}|$ mode peaks from NR fits, we can shift each mode by $\Delta t=t^{\rm p}-t^{\rm p}_{220}$, with $t^{\rm p}_{220}<t^{\rm p}_{lm0}$~\cite{forteza2020,Estelles:2020twz}.
This induces a correction -
\begin{equation}
\label{eq:amperror}
    A^R_{lmn}(t^{\text{p}})=A^R_{lmn}(t^{\text{p}}_{220})e^{\frac{\Delta t}{\tau_{220}}-\frac{\Delta t}{\tau_{lmn}}}\,.
\end{equation}
Notice that the $\tau_{lmn}$ are comparable for all the modes studied here. For example, for a BH with $a_f\approx0.85$ (consistent with GW190521~\cite{LIGOScientific:2020iuh,Capano:2021etf}) $\tau_{330}/\tau_{220} \sim 0.986$, $\tau_{440}/\tau_{220} \sim 0.974$, and $\tau_{210}/\tau_{220} \sim 0.956$.
Here a conservative choice ${\Delta t \approx 10M}$ translates to
a $\approx (1,2,3)\%$ correction for the amplitude of the ($3,3,0$), ($4,4,0$), ($2,1,0$) modes, respectively; this is well below the current statistical uncertainties~\cite{Capano:2021etf}. This conservative choice corresponds to $\Delta t\approx 2 (t^{\rm p}_{330}-t^{\rm p}_{220})\approx (t^{\rm p}_{210}-t^{\rm p}_{220})$~\cite{forteza2020} and so sets a conservative upper bound on $\Delta t$.

A similar correction needs to be accounted for the intrinsic phase, $\phi_{lmn}(t^{\rm p})=\phi_{lmn}(t^{\rm p}_{220})+\omega_{lmn} \Delta t$.
Using Eq.~\eqref{definitions}, this translates to
\begin{equation}
    \delta\phi_{lmn}(t^{\rm p})=\delta \phi_{lmn}(t^{\rm p}_{220})+ \left(\frac{m}{2} \omega_{220}-\omega_{lmn}\right)\Delta t. \label{dphaseshift}
\end{equation}
Since $\omega_{lmn}\approx \frac{l}{2} \omega_{220}$ for $l=m$ modes (note that this is an exact result in the eikonal $l=m\gg1$ limit~\cite{ferrari:1984zz,Cardoso:2008bp}), the phase correction induced by $\Delta t$ is $\approx{10\%}$ and $\approx{20\%}$ for the ($3,3,0$) and ($4,4,0$) mode respectively, when we assume $\Delta t = 10M$ (see Supplemental Material
). This is a conservative choice for the ($3,3,0$) mode as $\Delta t\lesssim 5 M$~\cite{forteza2020}; here we expect a systematic uncertainty no larger than $4\%$.
Therefore, for the $l=m$ modes, the NR phase fits can be compared to the measured posteriors inference as,
\begin{equation}
    \delta\phi_{lmn}(t^{\rm p})\approx\delta \phi_{lmn}(t^{\rm p}_{220}) \quad \forall\quad l=m.
\end{equation}
However note that for the ($2,1,0$) mode the last term in Eq.~\eqref{dphaseshift} yields a non-negligible uncertainty.

\noindent{{\bf{\em Application on GW190521.}}}
We exemplify our test on GW190521, the only GW event with reported subdominant angular mode in the ringdown~\cite{Capano:2021etf}. The total signal-to-noise ratio~(SNR) of this event is $\rho \approx 14$, of which $\rho\approx 12$ comes solely from the ringdown phase. This is a consequence of large total source mass of this binary system ($M_{\rm tot}=151^{+29}_{-17}M_\odot$~\cite{LIGOScientific:2020iuh} and the system is a convenient choice to demonstrate a proof-of-concept of the APC test.
\begin{figure}
\includegraphics[width=0.98\columnwidth]{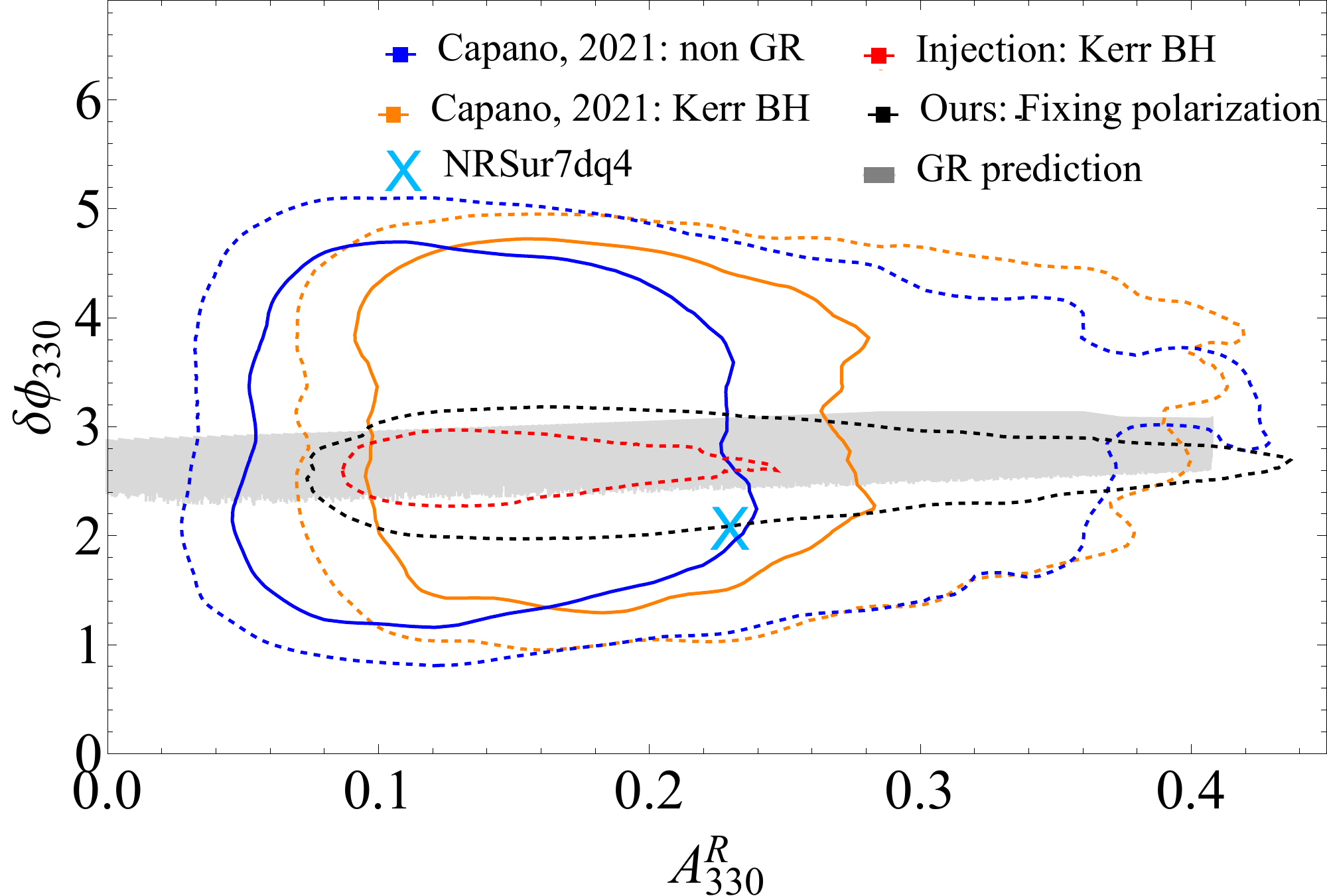}
    \caption{Ringdown APC test applied to GW190521. The amplitude ratio $A^R_{330}$ and the phase difference $\delta \phi_{330}$ obtained by various parameter estimations done on the data are compared to the GR allowed region in the $(A^R_{330},\delta \phi_{330})$ space (shaded gray area).
    The solid and dashed contours respectively represent the $67\%$ and $95\%$ crediblity contours obtained by Bayesian parameter estimation.
    The orange and blue contours correspond to the posteriors obtained by fixing the GR QNM spectrum and allowing deviations from the Kerr QNMs, respectively.
    The black dot-dashed contour provides the $95\%$ credible region obtained as in~\cite{Capano:2021etf} but fixing the polarization to the maximum likelihood value ($\psi=0$) given by the IMR analysis~\cite{Nitz:2020mga}.
    The red dot-dashed contour provides the $95\%$ credible region for the NR injection SXS:0258 consistent with GW190521 and with SNR $\rho=30$. 
    {The cyan cross corresponds to the $(A_{330}^{R}, \delta \phi_{330})$ fit to the maximum likelihood waveform obtained using the NRSur7dq4~\cite{Varma:2019csw} waveform approximant, which includes precession and binary parameters consistent to GW190521~\cite{Estelles:2021jnz}.}
    } 
    \label{fig:aratiophase}
\end{figure}

In Fig.~\ref{fig:aratiophase}, we first mark the GR-permissible region obtained by the NR fits as a grey band. Note that this region also accounts for the following uncertainties: the $1\sigma$ deviations on the best fit results, the $\sim 1\%$ and $\sim 4\%$ uncertainties on $A^R_{330}$ and $\delta\phi_{330}$ caused by ambiguity in ringdown start time, and the errors due to approximating the spheroidal harmonics as spherical harmonics. Next, we project the $67\%$ and $95\%$ credible regions of the measured posterior distribution for $A^R_{330}$ and $\delta \phi_{330}$ from ~\cite{Capano:2021etf} on to the $(A^R_{330},\delta \phi_{330})$ space. The posteriors obtained by assuming that ringdown has GR predicted Kerr QNM spectrum corresponds to the orange contour whereas the blue contour corresponds to the case where the QNM frequencies and damping times are allowed to vary freely. Interestingly, this more agnostic assumption does not deteriorate the confidence region significantly. 
The black dot-dashed contour provides the $95\%$ credible region obtained by fixing the polarization angle $\psi$ to the maximum likelihood value estimated from the full IMR analysis in~\cite{Nitz:2020mga}. This is similar to fixing right ascension and declination as done in~\cite{Capano:2021etf} (see \cite{Cotesta:2022pci,Isi:2022mhy,Finch:2022ynt} for a discussion on fixing these parameters in BH spectroscopy tests).
Estimating the polarization angle independently helps to break the degeneracy between $\psi$ and $\delta\phi_{lmn}$. $\psi$ can be estimated from the inspiral-merger regime, while the intrinsic dependence of $\delta\phi_{lmn}$ on $q$ and $\chiphen$ 
arises in the ringdown phase (see the Supplemental Material).

Lastly, we inject the NR waveform SXS:0258 into Gaussian noise at SNR $\rho=30$ for a 3-detector (LIGO-Hanford, LIGO-Livingston and Virgo) configuration and perform a parameters estimation using the PyCBC inference library~\cite{pycbc}. This numerical waveform has parameters compatible with GW190521~\cite{footnote}) but has been injected with twice the SNR of GW190521 to estimate the quality of the test achievable in the case of higher SNR events.  The red dot-dashed contour in Fig.~\ref{fig:aratiophase} denotes the $95\%$ credible region obtained for this case. We perform the parameter estimation at $t=t^p+15\rm ms$ (see the Supplemental Material).  We note that, as expected, the confidence region shrinks and the test is significantly more accurate with higher SNR.

We presented the main result of applying APC to GW19021 in Fig.~\ref{fig:aratiophase}; We find that the $1\sigma$ credible interval obtained in~\cite{Capano:2021etf} has a substantial support in the gray GR-permissible region marked on the $(A^{R}_{lmn}$, $\delta\phi_{lmn})$ space. Therefore, we conclude that the mode amplitude and phases measured in GW19021’s ringdown are compatible with the GR BBH predictions and this event passes the APC test.

\noindent{{\bf{\em Discussion.}}}
The APC test provides a novel strategy for testing GR using the ringdown mode excitations. NR waveforms of BBH mergers give accurate empirical relations between $A^R_{lmn}$ and $\delta\phi_{lmn}$ as functions of the binary’s mass ratio and spins. We found that only a narrow strip in the $(A^{R}_{lmn},\delta\phi_{lmn})$ space is admissible for ringdown modes of quasi-circular BBHs within GR. We build the APC test based on this feature and present a proof-of-concept implementation of this test on GW190521. \comment{We find that the $1\sigma$ posterior distributions of $A^{R}_{lmn}$ and $\delta\phi_{lmn}$ obtained in~\cite{Capano:2021etf} for this event has substantial support in the GR-permissible region, showing that GW190521 passes the APC test}.
\comment{Furthermore, we verify that the combined modelling uncertainties for the $(3,3,0)$ mode are well below the statistical uncertainties of the current GW observations. Overall, for the ($3,3,0$) mode the total systematic errors accumulated from the fit, ringdown start time, and spherical-harmonic approximation may reach a value $\sim 13\%\,$ for both $A^R_{lmn}$ and $\delta\phi_{lmn}$  (see Supplemental Material). For GW190521, the marginalized $1\sigma$ statistical uncertainties on these quantities are $ \sim 100\%$ — much larger than the systematic deviations accumulated from our fit uncertainties.  While this holds for any $l=m$ mode, the phase fits of $l \neq m$ modes are non-negligibly affected by the shift of the peak time; this makes $l\neq m$ modes not optimal for the proposed implementation of the test.}
%
%
%
However, the situation might change for louder detections as those routinely expected in the third-generation era~\cite{Maggiore:2019uih,Kalogera:2021bya}, in which case systematic errors of the fit might limit the accuracy of the APC test, unless the quality of NR waveforms improves. 
On the other hand, the measurements of the polarization angle $\psi$ (which is degenerate with the phase difference $\delta\phi_{lmn}$) and of other binary's intrinsic parameters are expected to improve as more interferometers are added to the network, or through an electromagnetic counterpart, and will anyway improve with third-generation detectors, therefore allowing for a more accurate test.



{We have focused on quasi-circular binaries with aligned spins, although we can extend a similar concept to build a more generic test. In the Supplemental Material
we show that the effect of eccentricity on the fits of $\delta\phi_{330}$ ($A^R_{330}$) is non-negligible only when $e\gtrsim 0.3$ ($e\gtrsim 0.6$). Therefore, the current implementation of the test is robust to mild eccentricities. By comparing the posterior distribution for $A^R_{330}$ shown in Fig.~\ref{fig:aratiophase} to the eccentricity fits, we have obtained a mild bound of $e\lesssim 0.9$ at the $95\%$ level on the eccentricity of GW19052. Note also that several works using the full IMR analysis on GW190521 have reported a moderately high effective precession spin parameter~\cite{schmidt:2014iyl}, $\chi_p=0.68^{+0.25}_{-0.37}$ (although waveform systematics and prior choices significantly affect the posterior estimates of this event~\cite{Estelles:2021jnz}).
Interestingly, when we fit for $\delta\phi_{330}$ and $A^R_{330}$ corresponding to the maximum likelihood waveform including precession~\cite{Varma:2019csw}, we notice that the effect of precession for GW190521 is within the measurement errors for this event. However, the best fit is marginally outside the grey shaded area (corresponding to the non-precessing scenario) in Fig.~\ref{fig:aratiophase}. Because of the large statistical error, the systematic effect of neglecting the spin precession does not affect GW190521 significantly. 
Therefore, we can use GW190521 as a proof of concept for the APC test, and future louder events could be used to constrain the binary precession independently from ringdown measurements only. A detailed examination of the effects of precession in the ringdown is an involved problem and requires a dedicated study.
}

\noindent{{\bf{\em Interpretation and  extensions.}}}
As with any null-hypothesis consistency test, its violation suggests a departure from the adopted baseline assumptions, and so a violation of the null-test could have various origins. We spell out the viable interpretation when an event does not pass the APC test — a)~Most conservatively, it might be evidence for mis-modelling the signal e.g., presence of {strong} spin precession or large eccentricity in the BBH; b)~It could be because the observed ringdown is not BH coalescence; note this does not preclude the remnant from being a standard Kerr BH. We expect the coalescence of massive neutron stars, boson stars~\cite{Palenzuela:2017kcg,Helfer:2018vtq,Bezares:2018qwa,Bezares:2022obu}, and other exotic compact objects~\cite{Cardoso:2019rvt} to produce QNMs consistent with Kerr BHs in GR. However, the QNM amplitudes and phases can be different from GR as the merger dynamics could be modified; these will therefore fail the APC test while being consistent with a traditional Kerr BH spectroscopy; c)~Finally and most radically, it could be because the underlying coalescence dynamics is not governed by GR. Disentangling these possibilities calls for a generalization of our fits to incorporate features like precession and eccentricity, louder ringdown detections, and detecting a population of them. For instance, if the violation of the test were due to not including eccentricity/precession in our fits, out of a population of ringdowns only a subgroup would be violating it. However, if GR dynamics were under question, there could be a ubiquitous violation of the test. 
In this context, although measurement errors are large, it is relevant that GW190521 passes the APC test. It would be interesting to assess whether this is in tension with alternative explanations for this event, e.g. a Proca star merger~\cite{Bustillo:2020syj}, by fitting the ringdown amplitudes and phases for Proca star merger waveforms and performing Bayesian model selection between the two hypotheses~\cite{Bustillo:2020syj}.


While we focused on the ringdown signal with prior knowledge of the binary’s extrinsic parameters (estimated either from the IMR analysis or from another independent sky localization), a variant of this test would be to estimate the initial binaries parameters with ringdown and check for consistency with IMR analysis. In principle one could invert the $A^{R}_{lmn}(q,\chi_{\rm pheno})$ and $\delta \phi_{lmn}(q,\chi_{\rm pheno})$ relations to infer an estimate of the mass ratio (and spins) from the QNM excitations. However, owing to the mild dependence of $\delta \phi_{lmn}$ on the binary parameters the quality of this test is expected to be rather poor. A more promising avenue is to neglect the phases and use only the amplitude ratios of several subleading QNMs. We discuss this in the Supplemental Material.
%
%
This is interesting for GW190521-like systems where the short signal duration and low SNR in the pre-merger part leads to controversial and model-dependent inference on the binary parameters ~\cite{Nitz:2020mga,Kastha:2021chr,LIGOScientific:2021djp,Estelles:2021jnz}. Also, higher sensitivity at low frequency (as expected for third-generation detectors) will improve this test significantly (see also \cite{Bhagwat:2021kwv} for a conceptual framework in this direction).
Overall, the APC test provides an excellent arena to complement standard BH spectroscopy tests in the strong-gravity regime, especially for the next-generation detectors.

\noindent{{\bf{\em Acknowledgments.}}}
We acknowledge the Max Planck Gesellschaft for support, and we are grateful to the Atlas cluster computing team at AEI Hannover for their help. The authors are specially thankful to Lionel London, Cecilio García-Quiros, and Juan Calderon-Bustillo for the invaluable discussions and further clarifications about the NR phase alignment and phase conventions. X. Jimenez is also thankful to P. Mourier for the useful discussions about the  correspondence of the fit and parameter-estimation results. S.B. is supported by the UKRI Stephen Hawking Fellowship, grant ref. EP/W005727.
P.P. acknowledges financial support provided under the European Union's H2020 ERC, Starting Grant agreement no.~DarkGRA--757480. We also acknowledge support under the MIUR PRIN (Grant 2020KR4KN2 ``String Theory as a bridge between Gauge Theories and Quantum Gravity'') and FARE (GW-NEXT, 
CUP:~B84I20000100001,  2020KR4KN2) programmes, and from the Amaldi Research Center funded by the MIUR program "Dipartimento di Eccellenza" (CUP:~B81I18001170001).
\bibliography{biblio.bib}
\appendix

\section{Fits for $A^{R}_{lmn}$ and $\delta \phi_{lmn}$}
\label{ap:NR_fits}
For each of the $\Nrcasesalln$NR SXS waveforms in our dataset, we fit for $A^{R}_{lmn}$ and $\delta \phi_{lmn}$ for $(l,m,n)=(3,3,0)$, $(2,1,0)$, and $(4,4,0)$ modes. The waveform dataset spans  $q\in\left[1,10\right]$ and $\chi_{1,2}\in\left[-0.9,0.9\right]$. 
We fit for the amplitudes $A_{lmn}$ and phases $\phi_{lmn}$ of each $h_{lmn}$ mode using the following ansatz,
\begin{equation}
\label{eq:rdh}
    h_{lmn}= A_{lmn}  e^{- i \phi_{lmn}} e^{i 2 \pi f_{lmn} t} e^{- t/ \tau_{lmn}}\,,
\end{equation}
where the frequencies and damping times are fixed to the values predicted by GR, while $A_{lmn}$ and $\phi_{lmn}$ are amplitudes and phases with same reference starting time $t=0$. However, the ringdown modes $h_{lmn}$ are extracted at a reference time $t^{\rm p}_{220}$ corresponding to the peak of the dominant ($2,2,0$) mode. To have the same reference starting time, we account for a time shift $\Delta t=t^{\rm p}-t^{\rm p}_{220}$, where $t^{\rm p}\lesssim10 M$ is the time at which the strain signal peaks (global peak).
As discussed in the main text, this ansatz does not fully match Eq.~\eqref{eq1} due to the peak time ambiguity. It adds an uncertainty on the phase values $\delta\phi_{lmn}$ for the $l=m$ modes but this is typically small.

In the rest of this appendix we provide a detailed discussion about the NR dataset used and the accuracy of the fits.

\subsection{Error estimate for NR waveforms}
NR waveforms contain two sources of uncertainties that are of interest to our study -- i)~resolution uncertainties which are produced by the finiteness of the numerical grid; and ii)~extrapolation errors which are produced from computing the data at a finite radii and extrapolating them to future null infinity. To quantify resolution errors, for each simulation we compute the mismatch (as defined, e.g., in Eq. (1) of~\cite{Bhagwat:2019dtm}) between waveforms at the two highest resolutions of the simulations in the SXS catalog. To estimate the extrapolation errors, we use the highest-resolution waveform and compute the mismatch between the waveform extrapolated with second or third polynomial order. We finally compute the distribution of the mismatch across our dataset. In Table~\ref{tab:NR-erros}, we give the order of magnitude of the mismatch at the median value for the distribution for each mode. We see that extrapolation errors are negligible relative to resolution errors for all modes, and the maximum mismatch is at most ${\cal O}(10^{-3})$. A more detailed study on NR systematics in the context of ringdown amplitude and phase fits will be presented in a companion paper~\cite{rdownampphase2}.
Next, we will also compare the results of our fits obtained using different NR BBH waveforms catalogs.

\begin{table}[]
\begin{tabular}{|c|c|c|}
\hline
\textbf{Mode} & \textbf{Error} & \textbf{Mismatch} \\ \hline
$h_{22}$      & Resolution     & $\mathcal{O}(10^{-3})$               \\ \cline{2-3} 
              & Extrapolation  & $\mathcal{O}(10^{-6})$               \\ \hline
$h_{33}$      & Resolution     & $\mathcal{O}(10^{-4})$               \\ \cline{2-3} 
              & Extrapolation  & $\mathcal{O}(10^{-6})$               \\ \hline
$h_{21}$      & Resolution     & $\mathcal{O}(10^{-3})$               \\ \cline{2-3} 
              & Extrapolation  & $\mathcal{O}(10^{-6})$                \\ \hline
$h_{44}$      & Resolution     & $\mathcal{O}(10^{-3})$               \\ \cline{2-3} 
              & Extrapolation  & $\mathcal{O}(10^{-6})$               \\ \hline
\end{tabular}
\caption{Resolution and extrapolation error estimates for NR waveforms in the SXS catalog used for our fits. }
\label{tab:NR-erros}
\end{table}

\subsection{Fits for $A^{R}_{lmn}$}
\label{ref:amp_fits}
For a quasi-circular nonprecessing binary, the mode amplitude ratios and phase differences generically depend on all intrinsic binary parameters such as mass ratio $q$ and spin amplitudes $\chi_{1,2}$.
However, it turns out that $A^{R}_{lmn}$ also depends more strongly on certain combinations of the spins, similarly to the effective spin parameters used in PN waveform modeling~\cite{borhanian:2019kxt}. 
We obtain analytical (approximate) relations for $A^{R}_{lmn}$ as a function of these BBH parameterizations using the ansatz~\cite{borhanian:2019kxt} 
\begin{align}
    \label{eq:amp_ans}
    A^R_{lm0}&=a_0 \,\delta + a_1\,\delta^2+ a_2\, \chi\,,  ~~~\forall \, \mathrm{odd~~modes}\\ 
    A^R_{lm0}&=a_0\,(1-3\eta)+a_1\,(1-3\eta)^2 +a_2\,(1-3\eta)^3 \\& a_3\, +\chi_s \,,\,\, \forall \, \mathrm{even~~modes}\nonumber
\end{align}
where the $\{(3,3,0), (2,1, 0 ) \} \in $
odd modes and $(4,4,0) \in$ even modes,  $\eta = q/(1+q)^2$, $\delta=\sqrt{1-4\eta}$, whereas
\begin{equation}
   \chi=\frac{\chi_a+\chi_s \sqrt{1-4\eta}}{2} 
\end{equation}
is the favoured combination of the spin parameters and, for (anti)aligned spins, $\chi_{s,a}= (m_1 \chi_{1}\pm m_2\chi_{2})/(m_1+m_2)$ with $m_{1,2}$ being the progenitor BH masses. 
This ansatz automatically enforces $ A^{R}_{lmn} (q)\to0$ for $q\to1$ for all odd modes in the $\chi_1=\chi_2$ limit, which arises from the binary’s symmetry under $m_1 \leftrightarrow m_2$. Note that the above ansatz differs from the ones we have used for the nonspinning fits presented in Ref.~\cite{forteza2020}. 

We fit the data in two hierarchical steps following~\cite{jimenez-forteza:2016oae}: we first fit the nonspinning waveforms using Eqs.~\eqref{eq:amp_ans} with $\chi_{1,2}=0$. Then, we fit for the spinning BBH waveforms, keeping the values of the coefficients obtained from the nonspinning fit to constrain the final result in the nonspinning limit. This improves the accuracy of the fit in the region of the parameter space where the NR simulations are known to be more accurate~\cite{jimenez-forteza:2016oae,keitel:2016krm}.
We get the following analytical ready-to-use fits:
\begin{align}
    A^R_{330}&=0.572 \sqrt{1-4 \eta }-0.144 (1-4 \eta )+0.035 \chi\,,
\end{align}
\begin{align}
    A^R_{210}&=\left| 0.328 \sqrt{1-4 \eta }+0.115 (1-4 \eta )-0.414 \chi \right|\,,
\end{align}
\begin{align}
    A^R_{440}&=0.251 \left(1+59.773 \eta ^3-16.307 \eta ^2-3 \eta \right)-0.011 \chi_s \,.
\end{align}

We choose to set the amplitudes to be positive by shifting phase by a factor $\pi$ i.e.,  $\delta\phi_{lmn}\rightarrow\delta\phi_{lmn}+\pi$, for those cases where the fit provides a negative amplitude.  For the ($2,1,0$) mode, we observe that $A^R_{210}$ tends to negative values at low mass ratio $q\lesssim 2$ and high spin $\chi\sim 0.4$. In this case we added the absolute value to the ansatz in order to keep the $A^R_{lmn}>0$ convention .
Note that the fits recover the test particle limit~\cite{forteza2020,Barausse:2011kb} as $q \rightarrow \infty$, and the nonspinning regime as $\chi_{1,2}\to0$. Indeed, in the nonspinning limit we verified that our fit agrees reasonably well with the fits in~\cite{forteza2020,ota:2019bzl,london:2014cma,london:2018gaq}.

\begin{figure}[t]
    \subfloat{\includegraphics[width=0.48\textwidth]{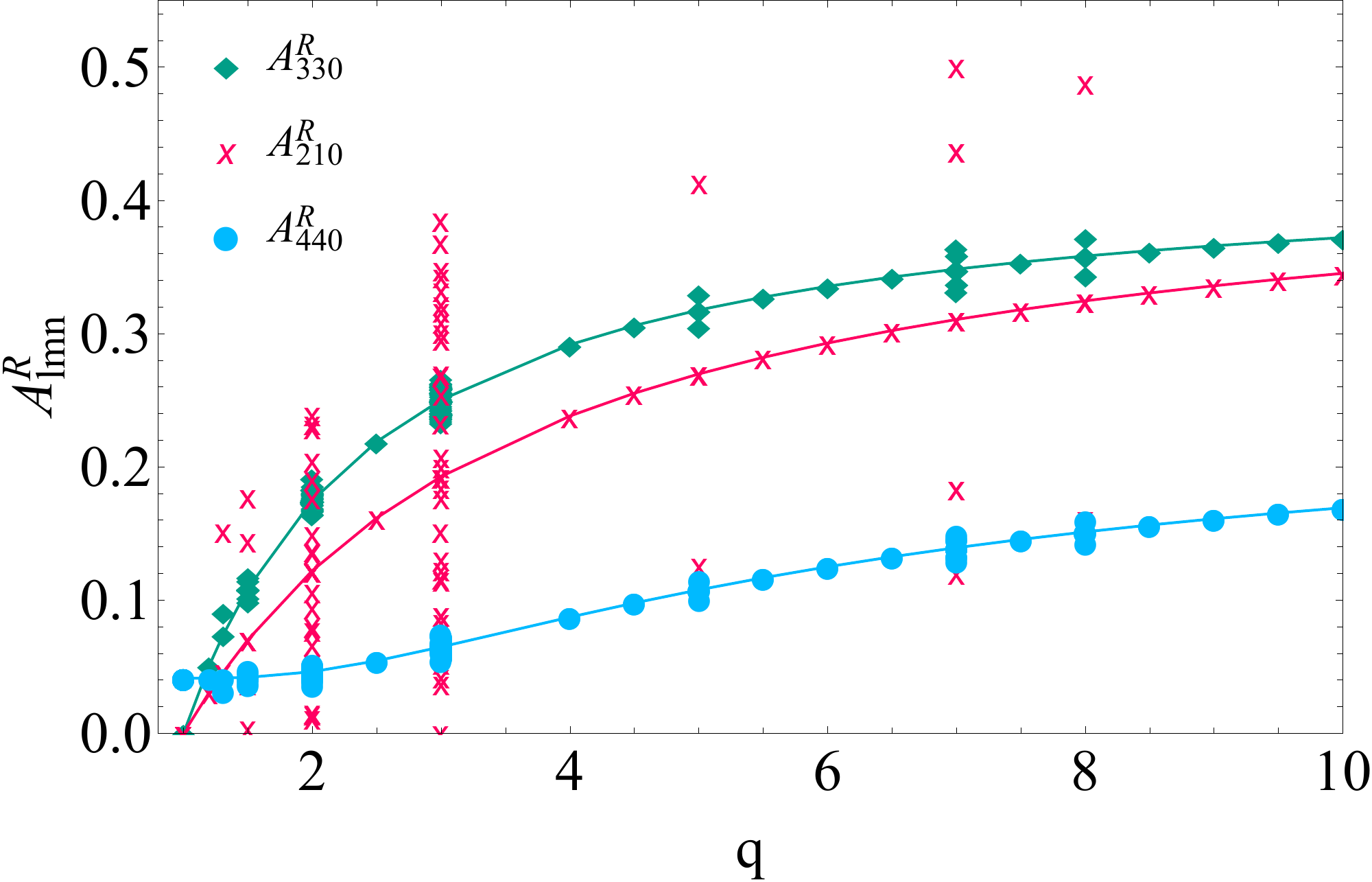}} \\
    \subfloat{\includegraphics[width=0.48\textwidth]{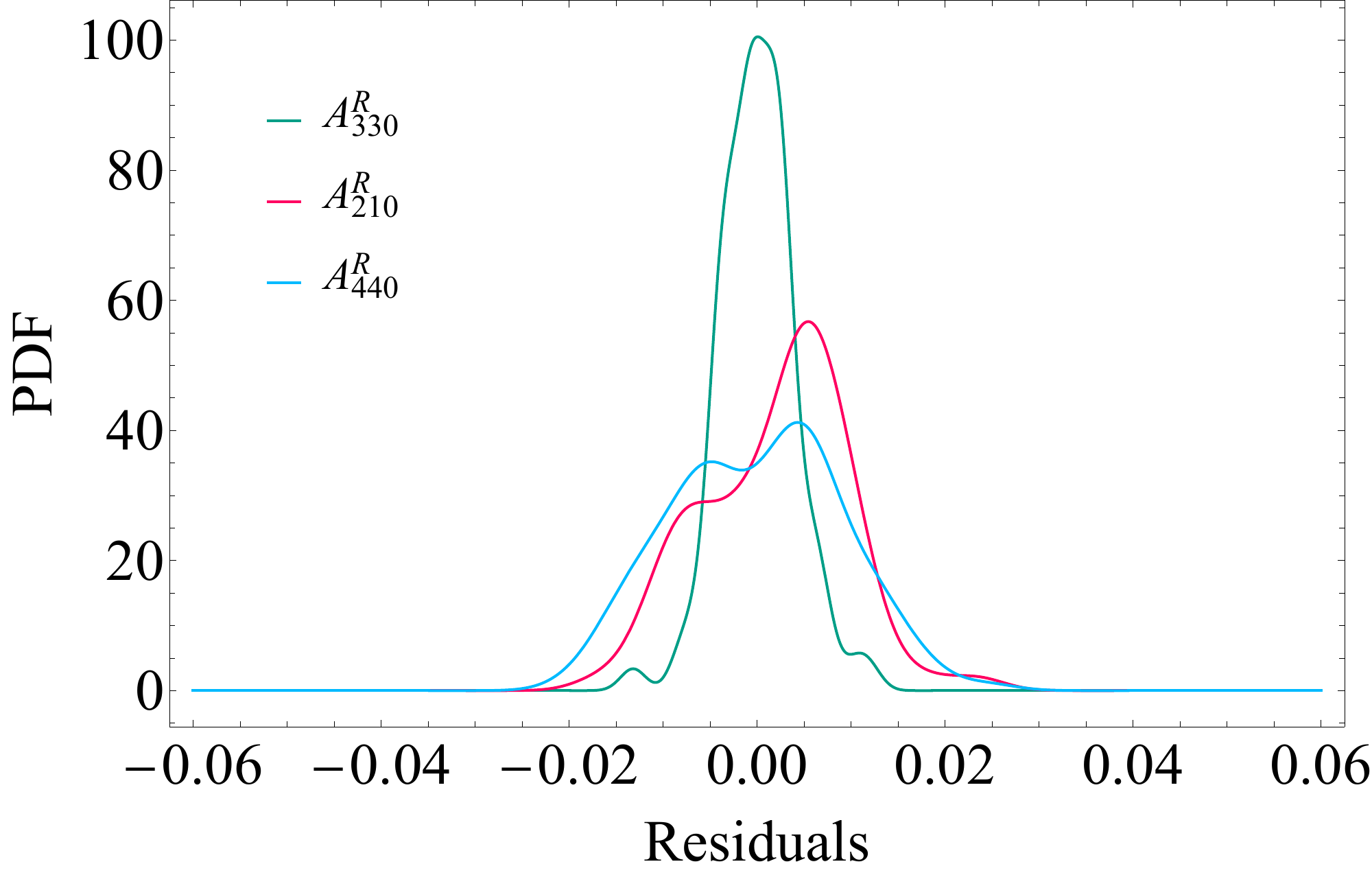}}\\ 
    \caption{Top panel: amplitude ratio $A^R_{lmn}$ in terms of the mass ratio $q$ obtained from fitting \Nrcasesalln waveforms for the (3,3,0), (2,1,0), and (4,4,0) modes. The spread of the points on the vertical direction quantifies the effects of the binary spin parameter, $\chi_{\rm pheno}=\chi,\chi_s$ for the odd/even modes, respectively. The solid line shows the fit for the nonspinning case. 
    Bottom panel: Normalised residual distributions for the three modes. Notice that effects are relatively small for the ($3,3,0$) and ($2,1,0$) compared to their typical amplitudes. Further analysis about the errors is provided in a followup paper~\cite{rdownampphase2}.
    }
    \label{fig:aratio_fits_q}
\end{figure}

In the top panel of Fig.~\ref{fig:aratio_fits_q}, we present the amplitude ratio $A^R_{lmn}$ as a function of $q$ for all \Nrcasesalln simulations. The green diamonds, red crosses, and blue dots correspond to $A^{R}_{330}$, $A^{R}_{210}$, and $A^{R}_{440}$, respectively. The solid lines denotes $A^R_{lmn}$ for the nonspinning BBHs, i.e., $A^{R}_{lmn} = A^{R}_{lmn}(q, \chi_{1,2} =0)$.
For all modes considere here, $A^R_{lmn}$ increases with the mass ratio, i.e. for more asymmetric binaries.
Spins effects are small for $A^{R}_{330}$ and $A^{R}_{440}$, leading to a small scatter around the solid lines.
This also suggests that spin effects are generically small for these modes, even when accounting for spin misalignment.

\begin{figure*}[]
     \subfloat{\includegraphics[width=0.32\textwidth]{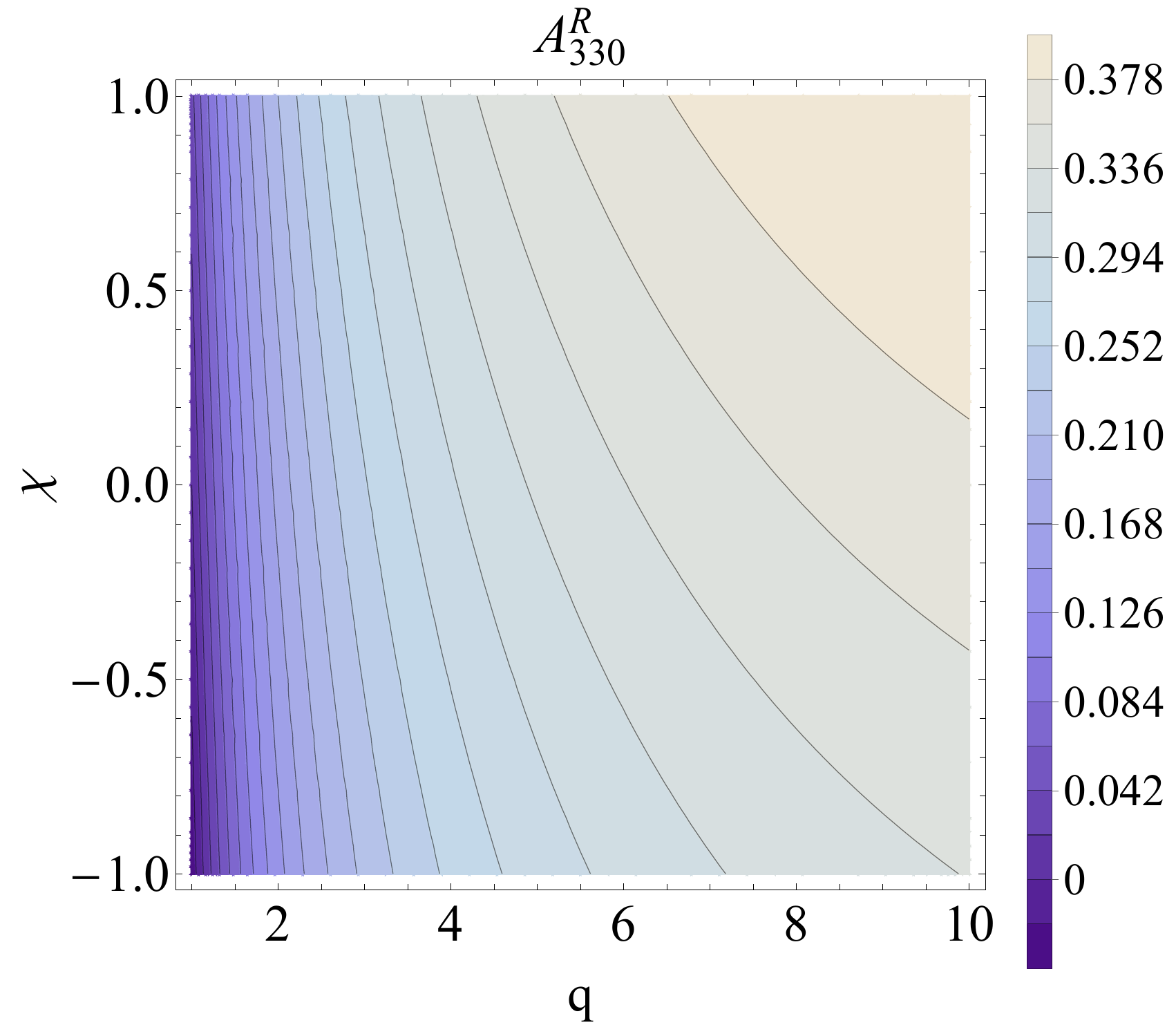}} 
     \subfloat{\includegraphics[width=0.32\textwidth]{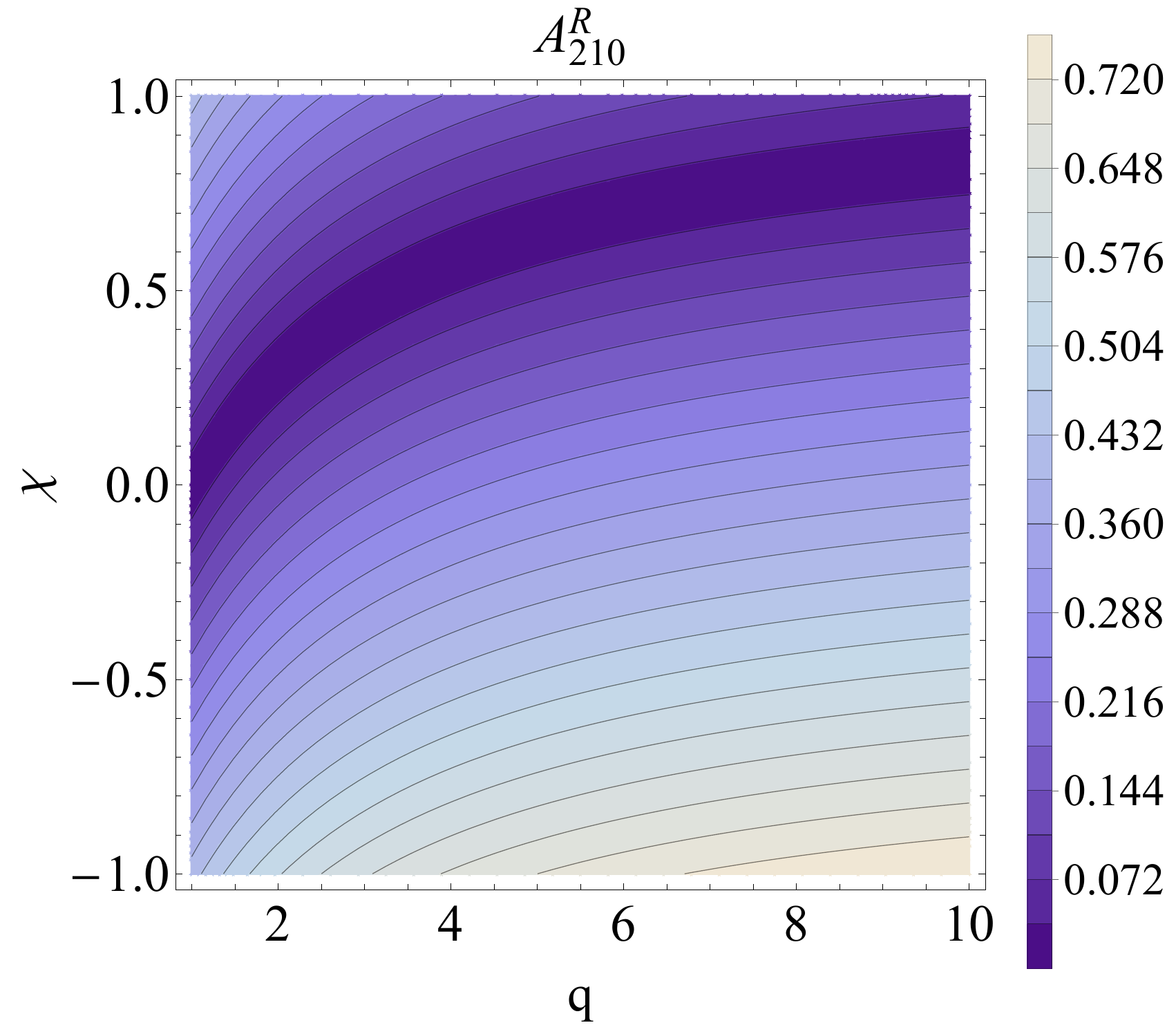}} 
     \subfloat{\includegraphics[width=0.32\textwidth]{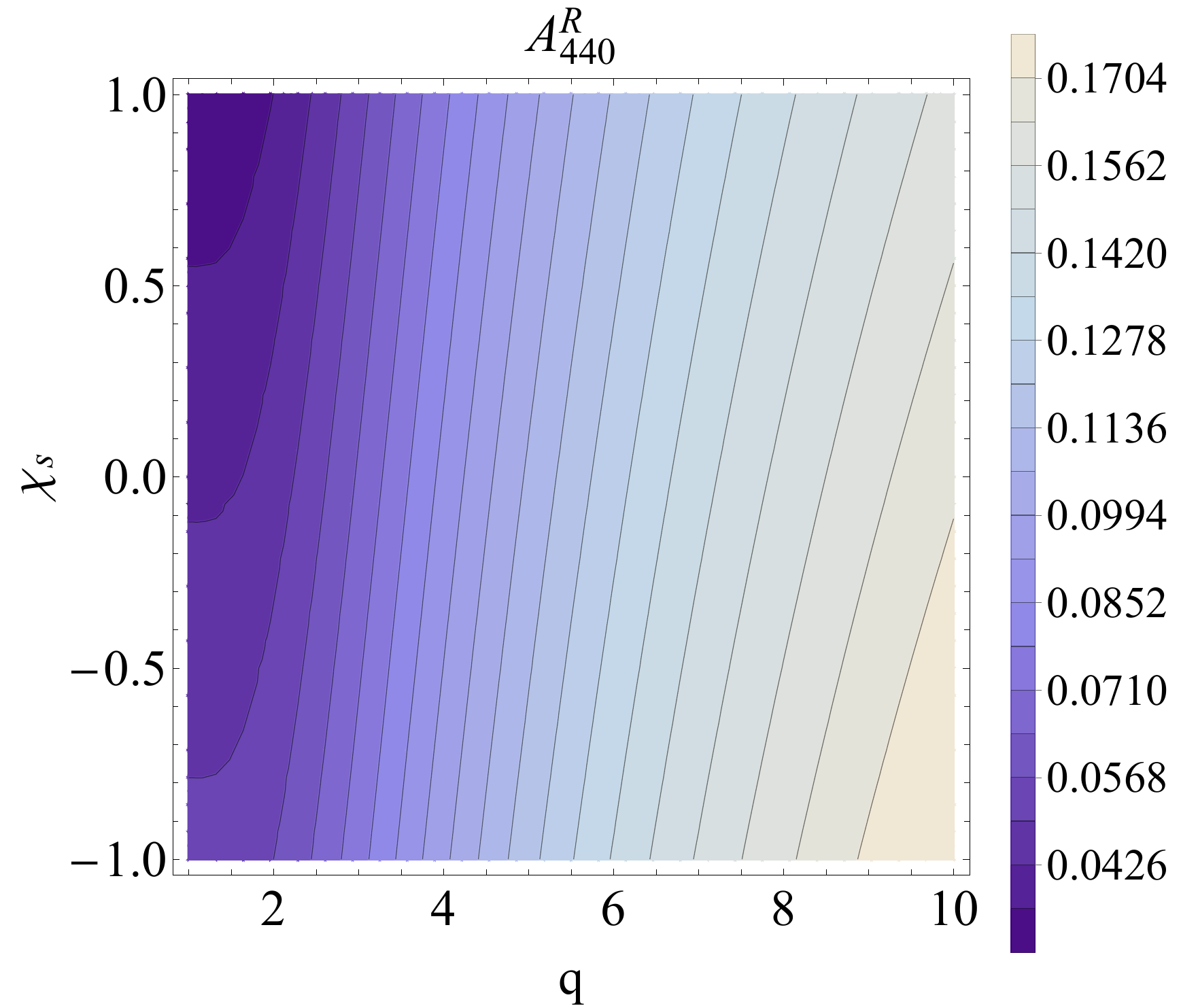}} \\
     \subfloat{\includegraphics[width=0.32\textwidth]{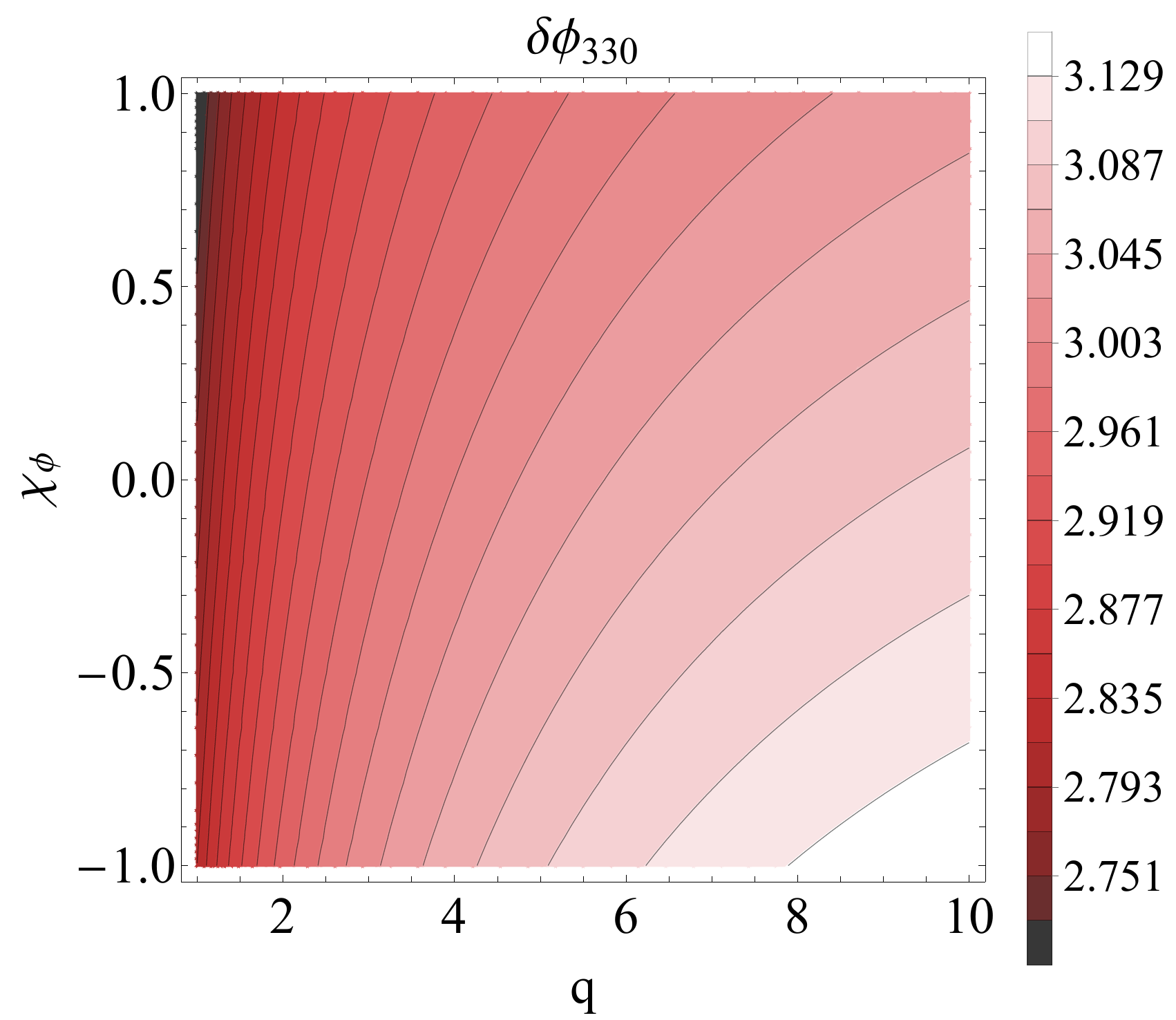}} 
     \subfloat{\includegraphics[width=0.32\textwidth]{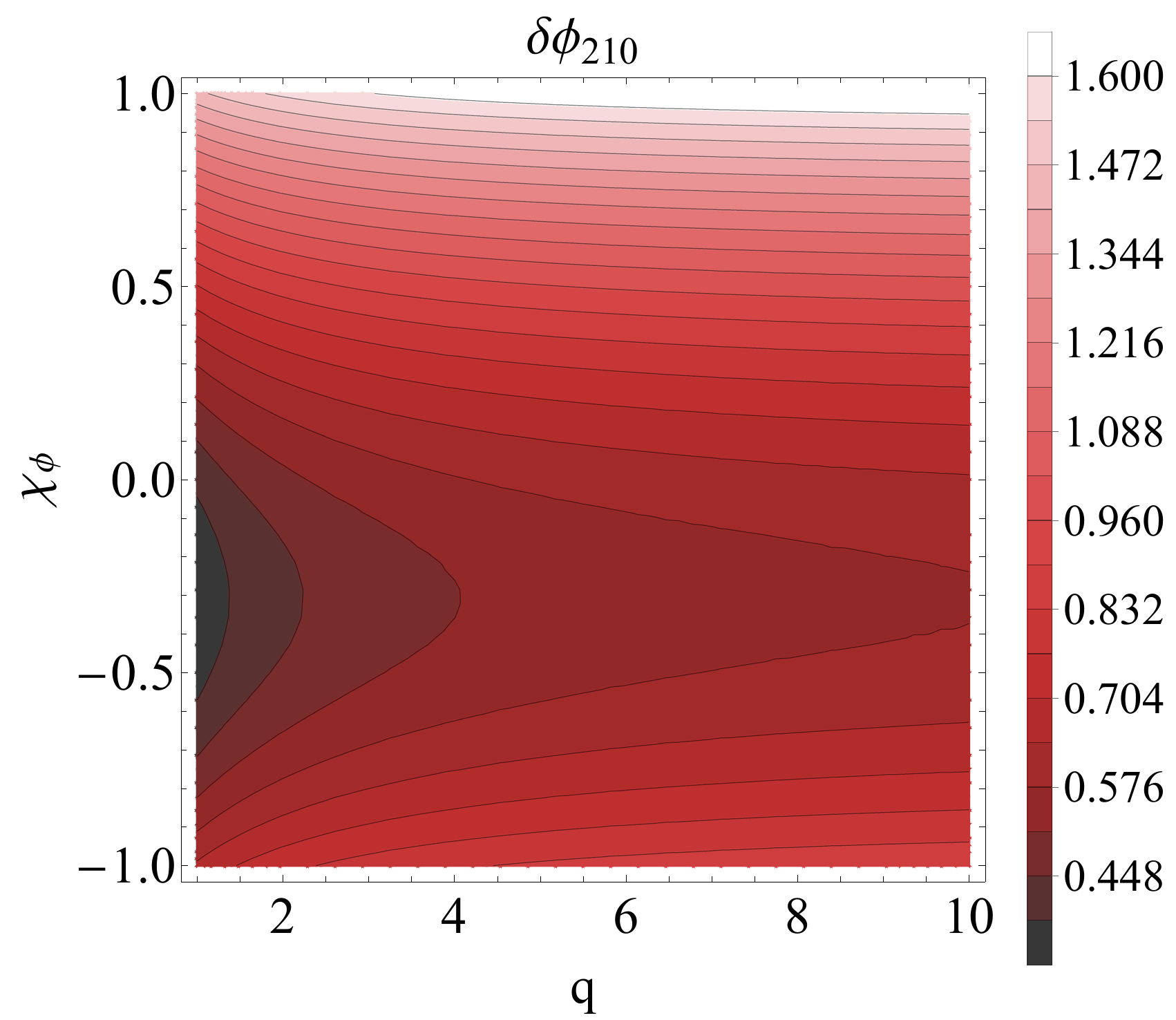}} 
     \subfloat{\includegraphics[width=0.32\textwidth]{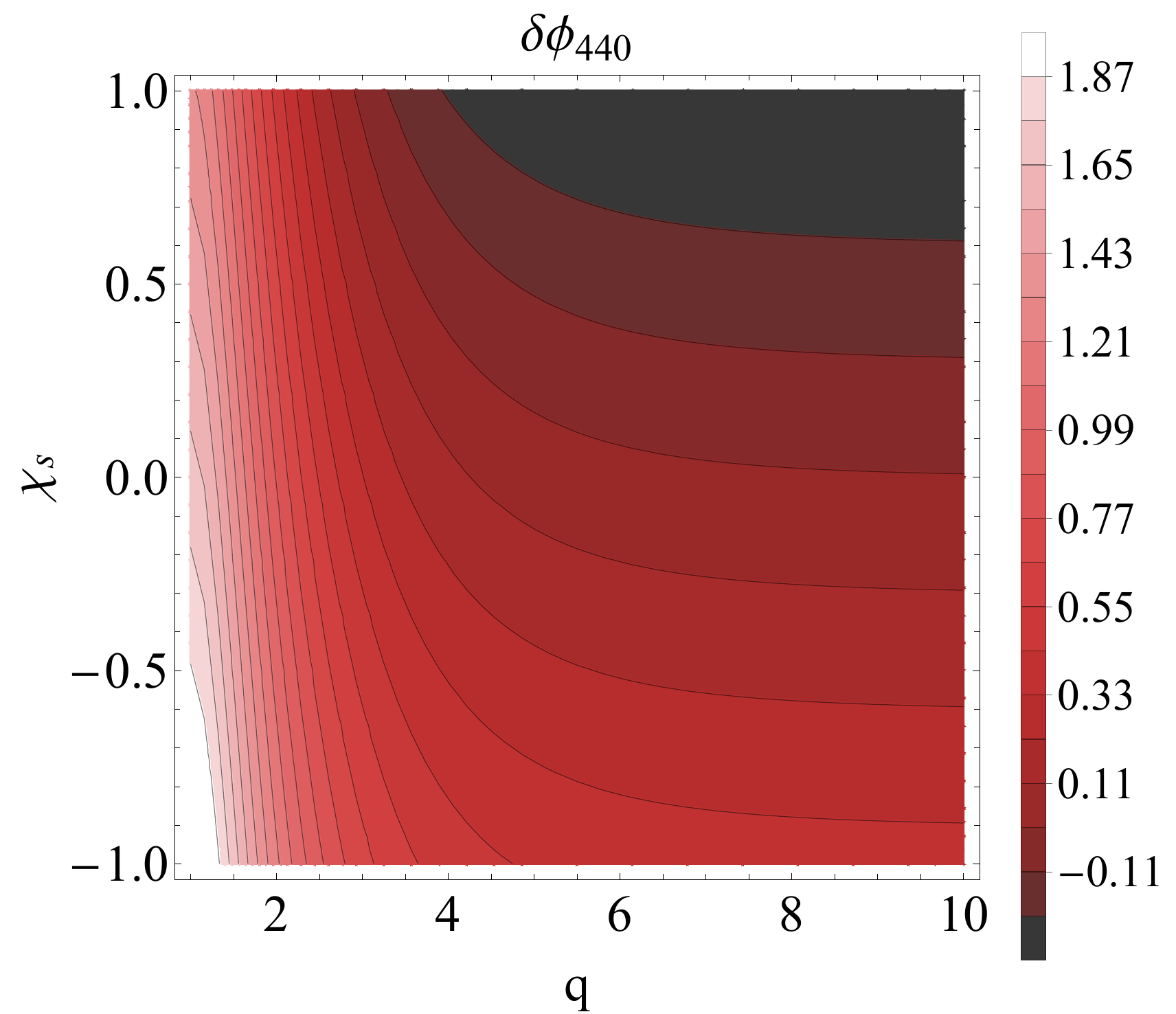}} 
    \caption{
    Contour plots showing the amplitude ratio $A^R_{lmn}$ (top) and phase difference $\delta\phi_{lmn}$ (bottom) as functions of $q$ and $\chi_{\rm pheno}=\chi,\chi_s,\chi_\phi$ (depending on the panels) for the $(l,m,n)=(3,3,0)$ (left), (2,1,0) (middle), and (4,4,0) (right) mode. As $\chi_{\rm pheno}$ increases, the magnitude of $A^R_{330}$ increases, while the  $A^R_{210}$ decreases and $A^R_{440}$ mildly decreases. The small region around $q\sim 1-2$ and $\chi>0$ in which $A^R_{210}$ increases with $\chi$ is forced by fitting $|A^R_{210}|$, which induces a sign flip on the trend of $A^R_{210}(\chi)$. 
    In the bottom panels, $\delta\phi_{lmn}$ decreases linearly with $\chi_\phi$ for the ($3,3,0$) mode while it increases as $\chi_\phi^2$ for the ($2,1,0$) mode. The value of the phase difference $\delta \phi_{lmn}$ decreases with $\chi_{\rm pheno}$ for the ($3,3,0$) and ($4,4,0$) modes while it increases for the ($2,1,0$) mode. } 
    \label{fig:contour}
\end{figure*}

In order to show the spin and mass-ratio dependence more clearly, in the top panel of Fig.~\ref{fig:contour} we present a contour plot of $A^{R}_{lmn}$ as a function of $q$ and $\chi,\chi_s$. We observe that, at variance with the other modes, $A^{R}_{210}$ depends significantly on the binary spins. 
Finally, we quantify the goodness of the fit using residuals. In the bottom panel of Fig.~\ref{fig:aratio_fits_q}, we present the normalized fit residual distributions, where residuals are the difference between the amplitude/phase obtained by the fit and that obtained from the raw NR data, i.e., ${\rm Residuals} =  (A^R_{lmn, {\rm NR-data}} - A^R_{lmn,  {\rm fit}}) $ or $ {(\delta \phi}_{lmn, {\rm NR-data}} - \delta \phi_{lmn, {\rm fit}})$ for each simulation used. We see that the residuals are centered around zero with a small spread. Compared to their absolute value, these errors are small for the (3,3,0) and (2,1,0) modes and modest for the (4,4,0) mode. 
\subsection{Fits for $\delta \phi_{lmn}$}
\subsubsection{Disentangling NR conventions from physical phase}
\label{subsec:conventions}
The NR waveforms from different catalogs adopt different conventions for phases, so one must appropriately account for this to combine/compare phases across several NR waveform catalogs. These conventions arise from the choice of -- a) the tetrad adopted to extract the NR waveform which adds an overall polarization angle $\psi_0$; and b) from rotations of the BH orbital plane by an angle $\varphi_0$. Two waveforms (say A and B) from different catalogs with the same physical intrinsic parameters and aligned in time are related by~\cite{CalderonBustillo:2015lrg,Garcia-Quiros:2020qpx},
\begin{equation}
h^{A}_{lmn}(t)=e^{\iota (\psi_0+ m\varphi_0)}h^{B}_{lmn}(t).
\end{equation}

The polarization angle across the NR codes is either $\psi_0=0$ or $\psi_0=\pi$ to preserve the rotating-counterrotating mode symmetry, $h_{lm}=(-1)^l h^{*}_{l-m}$, for circularly-polarized, nonprecessing waveforms\footnote{The rotating-counterrotating mode symmetry  implies $h_{lm}=(-1)^l\,h^*_{l,-m}$. We can define a new waveform $h^p_{lm}$ up to a polarization angle  $\psi_0$ as $h^p_{lm}=e^{i \psi_0}(-1)^l\,h_{lm}$ and $h^p_{l-m}=e^{i \psi_0}(-1)^l\,h_{l-m}$. Then, $h^p_{l-m}=e^{2 i \psi_0}(-1)^l\,h^{p,*}_{lm}$ and $h^p_{l-m}=(-1)^l\, h^{p,*}_{lm}$ if and only if $\psi_0=0,\pi$.}~\cite{Garcia-Quiros:2020qpx}. We are interested only in the physical contribution to the phase, $\delta_{lmn}(\vec{\lambda})$, which depends only on the binary intrinsic parameters $\vec\lambda$. A generic ringdown phase $\phi_{lmn}$ results from the sum of the three contributions~\cite{CalderonBustillo:2015lrg,london:2014cma,london:2018gaq}
\begin{equation}
\label{eq:phase_diff_all}
\phi_{lmn}=\delta_{lmn}(\vec{\lambda}) +m \,\varphi_{0}+\left\lbrace 0,\pi\right\rbrace\,.
\end{equation}
Unlike the physical phase $\delta_{lmn}(\vec{\lambda})$ that depends on the BBH parameters, the extrinsic phase terms $\psi_0$ and $\varphi_0$ may vary across sets of NR simulations and codes. However, the dependence on $\varphi_0$ is eliminated out if we fit for the following quantity
\begin{align}
\label{eq:phase_difference}
\delta \phi_{lmn}&\coloneqq\frac{m}{2}\phi_{22n}-\phi_{lmn}=\frac{m}{2}\delta_{22n}(\vec{\lambda})-\delta_{lmn}(\vec{\lambda})\\&+\left\lbrace 0,\left(\frac{m}{2}-1\right)\pi\right\rbrace \nonumber \,.
\end{align}
The phase difference $\delta \phi_{lmn}$ depends only on the intrinsic binary parameters $\vec{\lambda}$ and on a global phase factor which is either zero or $(\frac{m}{2}-1)\pi$ depending on the simulation. For instance, we can identify the convention used in a NR waveform by knowing that, in the low-frequency inspiral  regime, the phase difference between the dominant $(2,2,0)$ mode and a higher $(l,m,n)$ mode satisfies $\text{mod}(m\,\phi_{220}-2\phi_{lmn},2\,\pi)=0,2\pi$, for both the even and the odd modes (see Appendix D of~\cite{Garcia-Quiros:2020qpx} and~\cite{Estelles:2020twz}). For the SXS data, we have checked that this value is consistent with $\psi_0=0$. The RIT and Maya~\cite{ritcatalog,gatechcatalog} waveform catalogs (considered later on) adopt instead the $\psi_0=\pi$ convention~\cite{sxscatalog}.

\subsubsection{Phase fits}

Similar to the case for the amplitude ratio, we produce ready-to-use fits for $\delta \phi_{lmn} = \frac{m}{2}\phi_{22n}-\phi_{lmn}$ as a function of the BBH parameters. We use the following ansatz informed by the leading order PN expressions on $\delta \phi_{lmn}$~\cite{cotesta:2018fcv}
\begin{align}
    \delta\phi_{330}&= b_0\,\delta+b_1\, \chi_\phi +c_0\,,\\
    \delta\phi_{210}&= b_0\, \delta +b_1 \,\chi_\phi+b_2\, \chi_\phi ^2+c_0\,,\\
    \delta\phi_{440}&=a_0 \,\eta^{d_0}+b_1\, \chi_s\,,
       \label{eq:phase_ans}
\end{align}
where $\chi_\phi \equiv \frac{1}{2}(\chi_a+\frac{\sqrt{1-4\eta}}{1-2\eta}\chi_s)$ is another phenomenological fit parameter.

We follow a similar hierarchical fitting procedure as the one previously described for the amplitude-ratio fits. The fits obtained for the $\delta \phi_{lmn}$ read
\begin{align}
    \delta\phi_{330}&=2.759+0.406 \sqrt{1-4 \eta }-0.055 \chi_\phi   \\
    \delta\phi_{210}&=0.401+0.286 \sqrt{1-4 \eta }+0.402 \chi_\phi +0.652 \chi_\phi ^2 \nonumber\\
    \delta\phi_{440}&=4245.459 \eta ^{5.646}-0.365 \chi_s \nonumber  \,.
\end{align}

\begin{figure}[h]
    \subfloat{\includegraphics[width=0.48\textwidth]{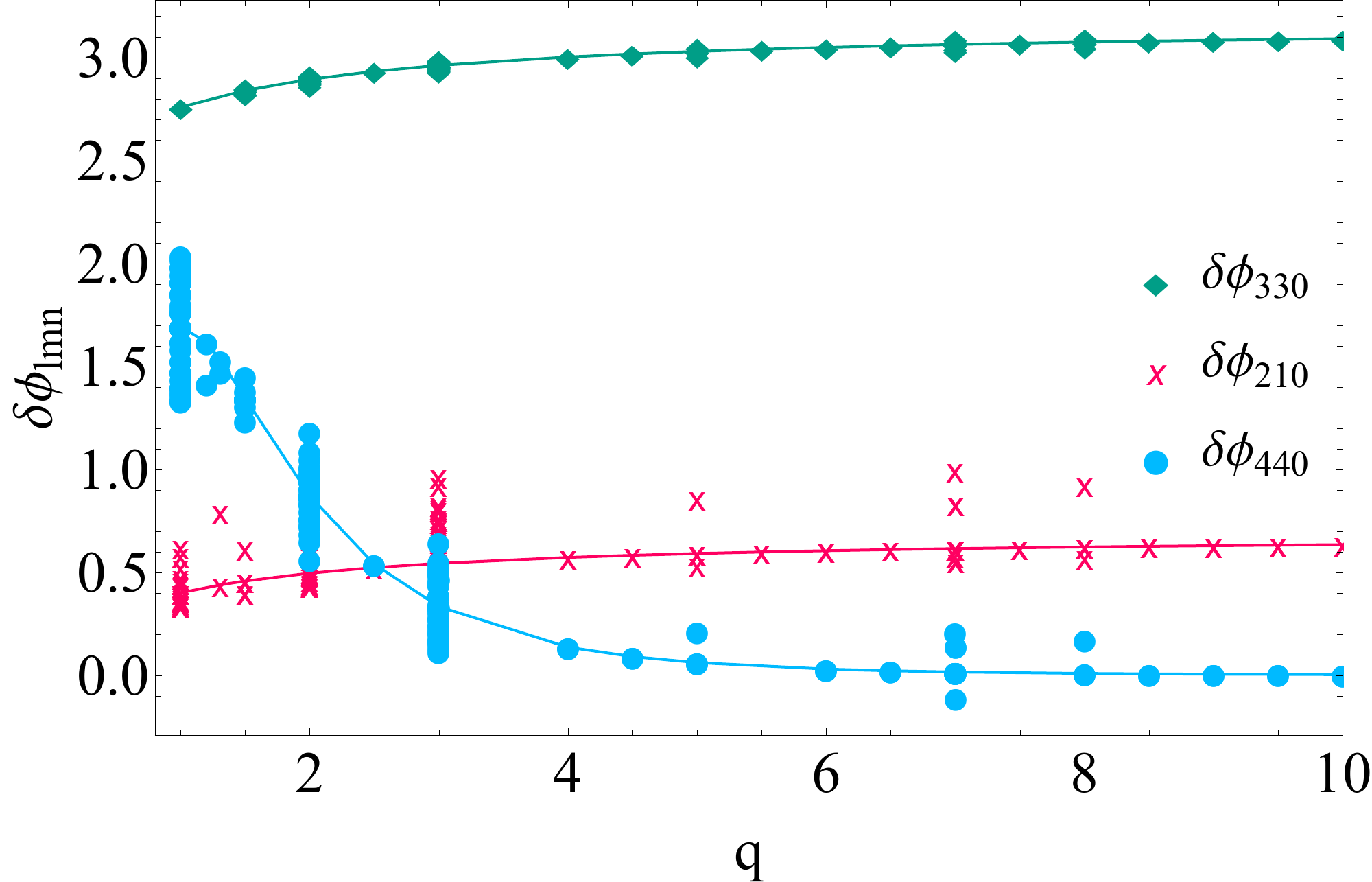}} \\
    \subfloat{\includegraphics[width=0.48\textwidth]{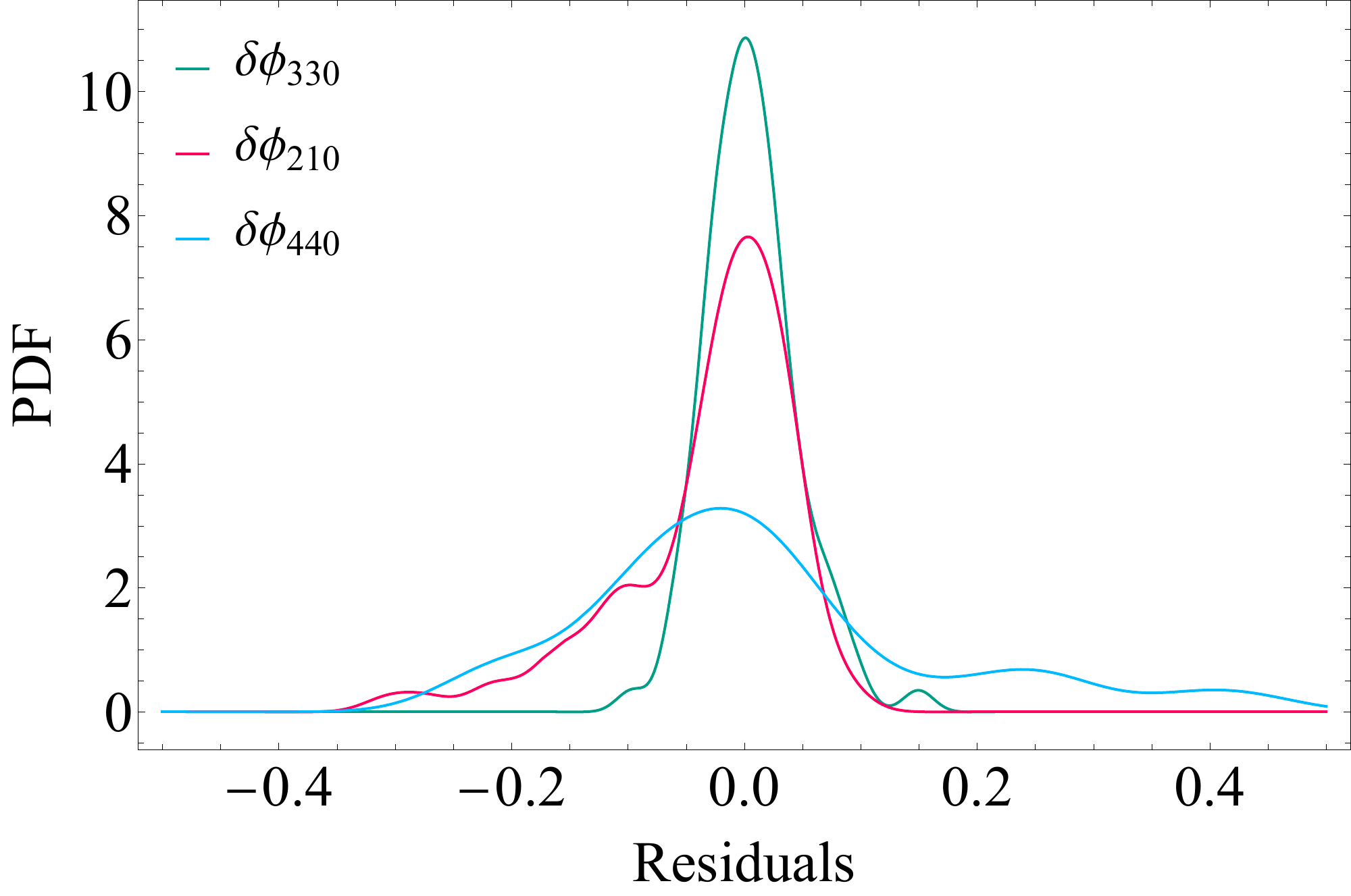}}\\ 
    \caption{Same as Fig.~\ref{fig:aratio_fits_q} but for the phase difference $\delta\phi_{lmn}$.}
    \label{fig:phase_fits_q}
\end{figure}

In Fig.~\ref{fig:phase_fits_q} we present $\delta \phi_{lmn}$ (as defined in Eq.~\eqref{eq:phase_difference}) obtained at $t_r-t^{\rm p}_{22}=10M$.
Again the solid lines correspond to the nonspinning BBH case, while the spread of points around the lines quantifies the spin dependence of the result. {As $q$ increases, we observe that $\delta \phi_{330}$ and $\delta \phi_{210}$ mildly increase while $\delta \phi_{440}$ rapidly vanishes.}
We stress that for the SXS catalog the convention is such that $\text{mod}(\delta \phi_{330},\pi)=\pi$ in the low-frequency PN regime.   
%
%
For $q\approx1$, we see that $\delta \phi_{lmn}$  develops a dependence on the BBH spin parameters, which is reduced as $q$ increases. This is consistent with~\cite{cotesta:2018fcv}, where $\text{mod}(\delta \phi_{330},\pi)$ is  evaluated at $t_r=t^{\rm p}_{22}$. 
We again evaluate the fit residuals for each NR waveform; the distribution of the residuals is shown in the bottom panel of Fig.~\ref{fig:phase_fits_q}.  Similar to Fig.~\ref{fig:aratio_fits_q}, the largest errors are obtained for $\delta \phi_{440}$. Finally, in the bottom panel of Fig~\ref{fig:contour} we show  $\delta\phi_{lmn}$ as a function of $q$ and $\chi,\chi_s$, $\chi_\phi$ for the ($3,3,0$) (left), ($2,1,0$) (middle) and ($4,4,0$) (right) modes. As in the case for the amplitude ratio, the phase $\delta\phi_{330}$ is only mildly affected by the spin, whereas a stronger spin dependence occurs for the (2,1,0) and ($4,4,0$) modes.  \comment{Requiring the amplitude of the (2,1,0) mode to be non-negative makes it a non-smooth function in the 
spin parameter space $\chi$~\cite{Garcia-Quiros:2020qpx} and induces the sudden increasing of $|A^R_{210}|$ at $q\sim 1-2$ and $\chi>0$. This originates the small bi-modal bump observed in the shaded area of the (2,1,0) mode of Fig.~1.}

\subsection{Comparison of the fits using other NR catalogs}
\label{ap:catalogs}
\begin{figure*}[]
    \includegraphics[width=0.48\textwidth]{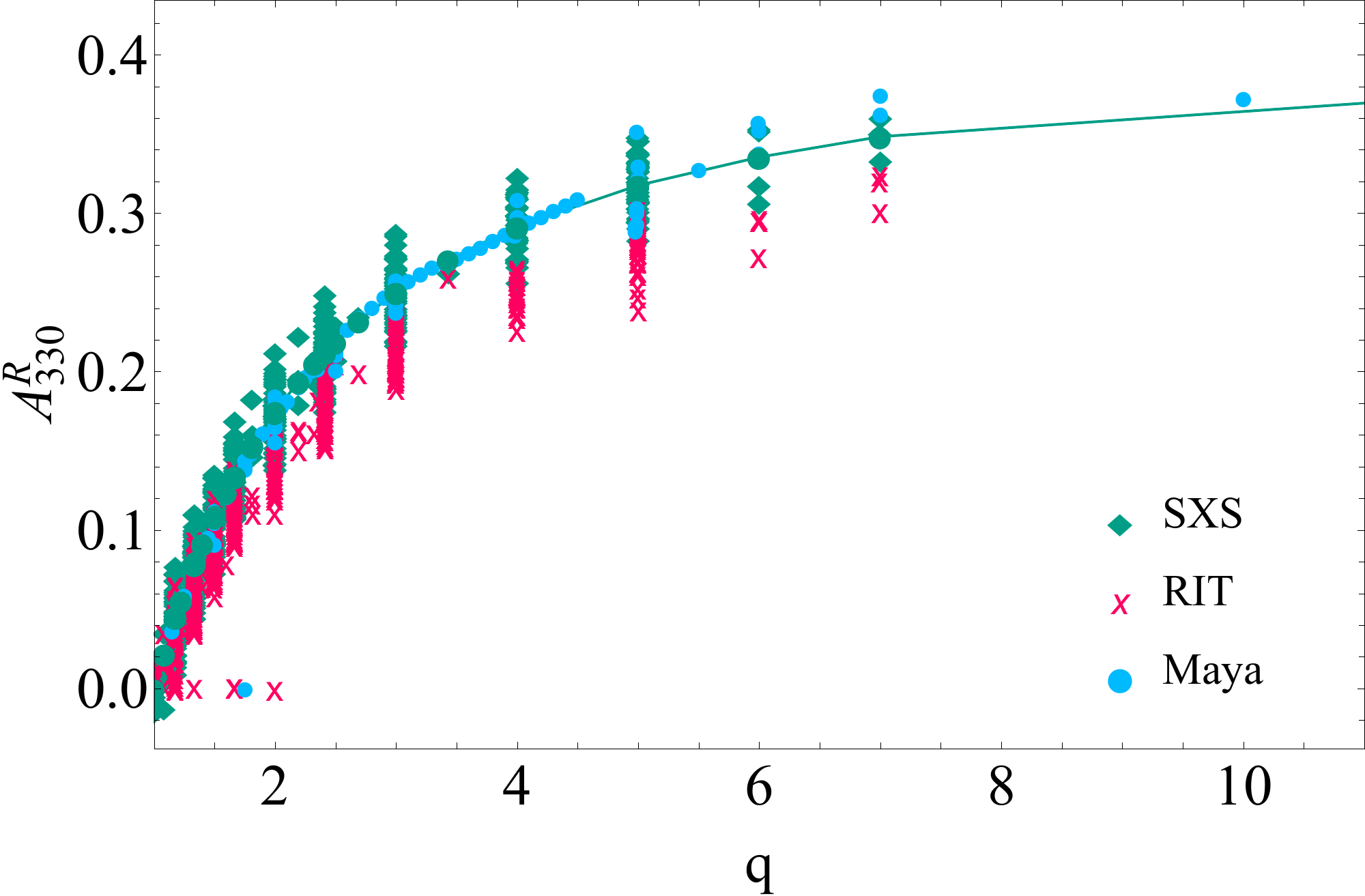}
    \includegraphics[width=0.48\textwidth]{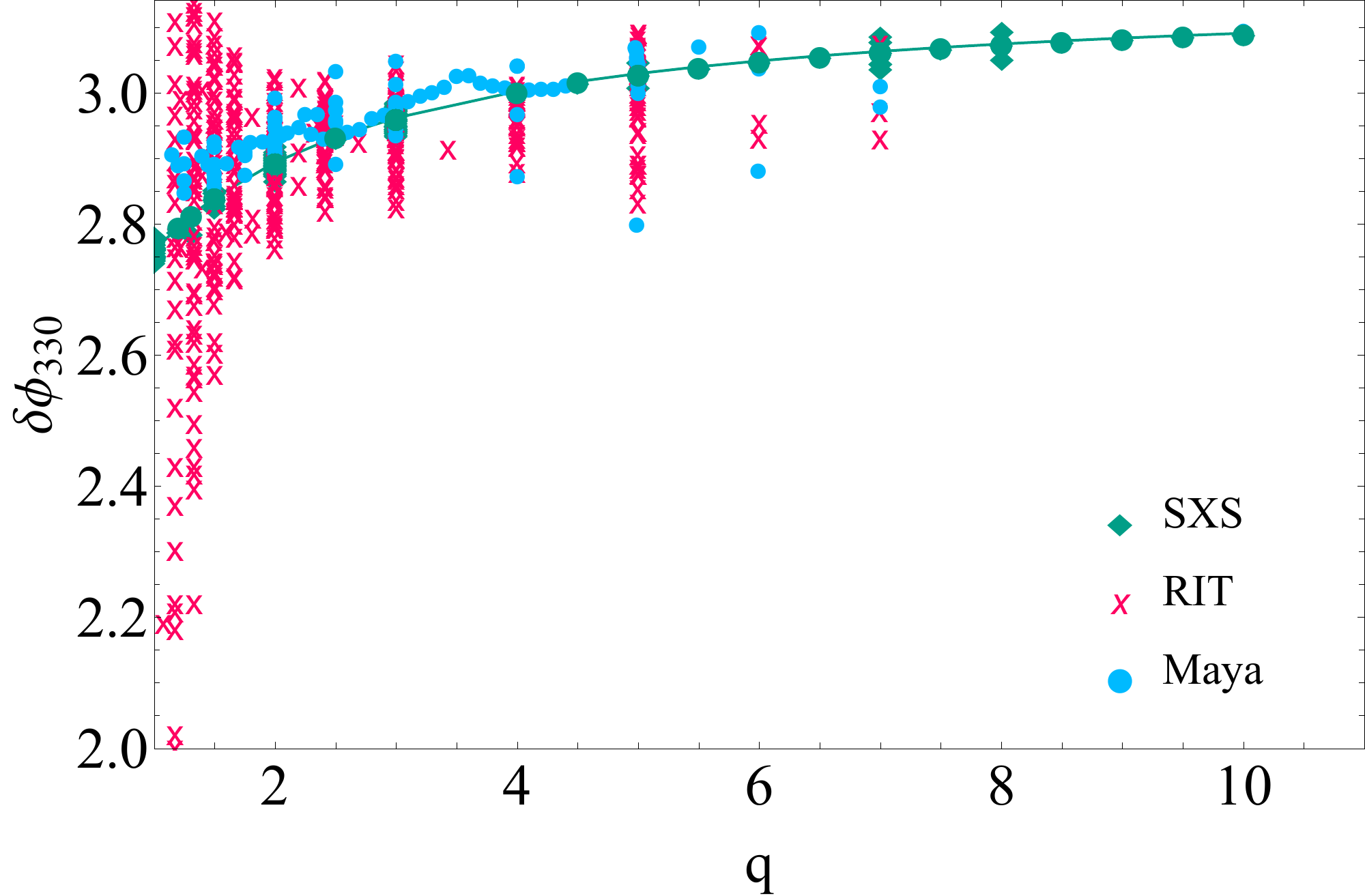}\\
    \includegraphics[width=0.48\textwidth]{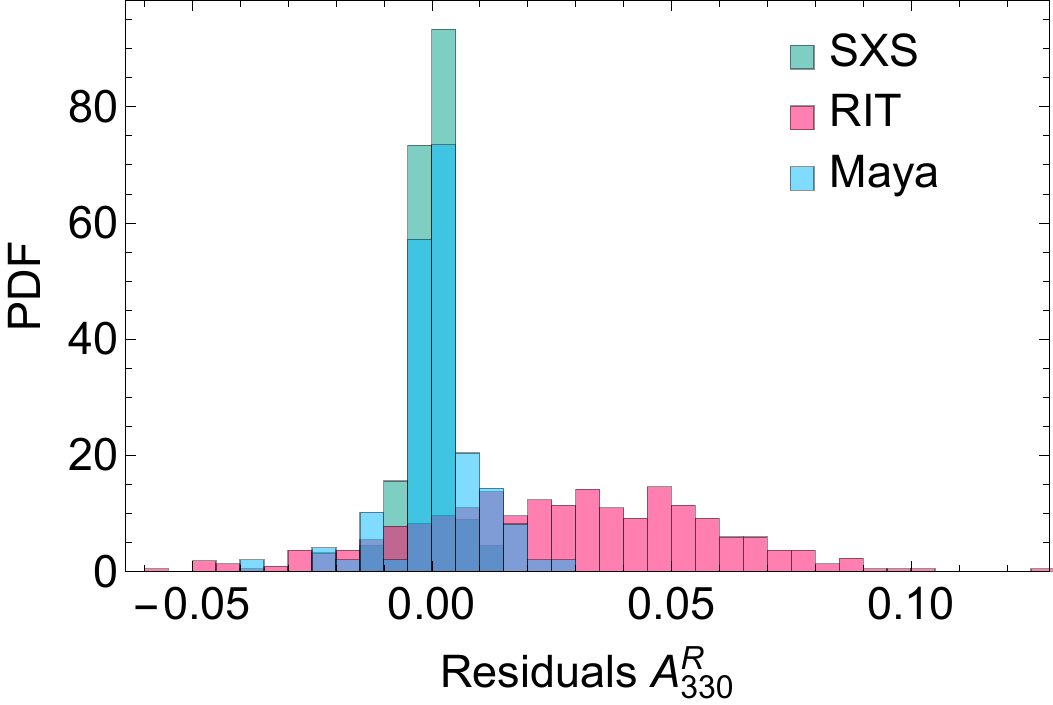}
    \includegraphics[width=0.48\textwidth]{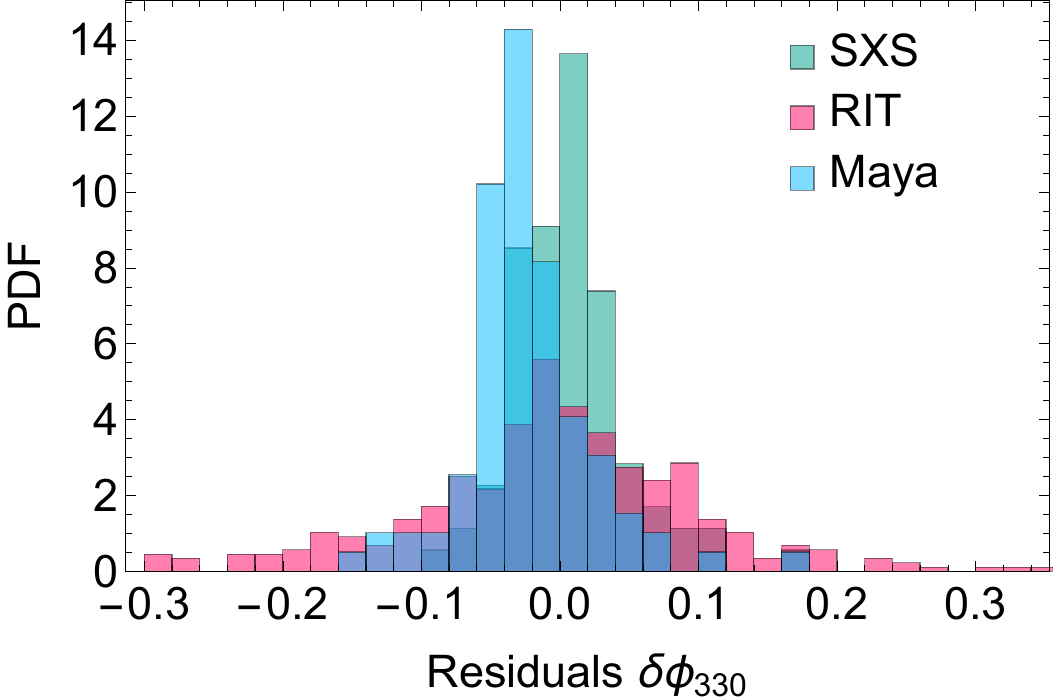}
    \caption{Top panel: Amplitude ratio (left) and phase difference (right) in terms of the mass ratio $q$ and the phenomenological spin parameter $\chi_{\rm pheno}$ obtained by fitting \Nrcasesalln waveforms for the (3,3,0) modes. Data from the SXS, RIT, and Maya catalogs are used here and the denoted by green, pink and blue respectively. \comment{The solid line joining the points represents nonspinning simulations using just the SXS catalog}. Bottom panel: Normalized residual distributions obtained for the three catalogs for the amplitude ratio (left) and phase difference (right).}
    \label{fig:amp_fits_q_codes}
\end{figure*}
The RIT and Maya public catalogs~\cite{ritcatalog,gatechcatalog} provide a large set of NR simulations that can be used for fitting and testing. In particular, the public data provided for the RIT catalog is tested to be in the convergent regime -- resolution errors shall dominate -- and shows good global IMR agreement with the SXS data for all the modes up to $l=5$. We have calibrated  the amplitudes and phases using data from the SXS catalog since they provides data at different resolutions and extrapolation levels. We use these for the error estimates previously presented in this study. However, we can use NR data from the RIT and Maya catalogs to benchmark our results. This test is particularly useful since the two families of codes use significantly different numerical schemes to solve Einstein's equations for a BBH~\footnote{For instance, while the SXS waveforms are solved in the generalized harmonic gauge, the RIT and Maya are solved using the BSSNOK formulation.}. We see that both $A^R_{lmn}$ and $\delta\phi_{lmn}$ computed from different codes are affected dominantly by the finite extraction and extrapolation effects -- adding up the differences between them to a few percent~\cite{keitel:2016krm}. We observe that the systematic errors (i.e., a shift of the median value of the distribution with respect to zero), are below the fitting errors characterized by the width of the distributions. This observation holds true for all the modes considered here.

To compare the data from the different catalogs, we first revisit the various conventions used in each of them -- for example, the $\psi_0=0,\pi$ rotational factors arising from different tetrad choices in the simulations. 
For the RIT and Maya catalogues, we need to replace $\delta\phi_{lmn}\rightarrow -\delta\phi_{lmn}-\frac{m}{2}\pi$, This factor results from the different tetrad conventions used in these codes. The additional minus sign comes from the reversed definition of the imaginary component of the $h_{lm}$ modes between the SXS dataset and the RIT and Maya datasets, which just implies that $\phi_{lmn}^{\text{SXS}} \leftrightarrow -\phi_{lmn}^{\text{RIT},\text{Maya}}$\cite{sxscatalog}. 

In Fig.~\ref{fig:amp_fits_q_codes}, we compare amplitude ratio and phase difference of the ($3,3,0$) mode obtained using SXS simulation to that given by the RIT and Maya wavefroms. 
For the amplitude ratio (left panels of Fig.~\ref{fig:amp_fits_q_codes}), we find a good match between the Maya and SXS data while the results show an offset of about $A^{R}_{330}\sim0.05$ when using the waveforms from the RIT catalog. This systematic offset is still smaller than the value of the uncertainty observed in GW190521, $\delta A^R_{330}\sim 0.1$, but will become important for louder events in the future. Regarding the phase difference (right panels of Fig.~\ref{fig:amp_fits_q_codes}), despite we find a resonably good agreement between all the three codes, the RIT code shows a slightly larger tails. 
The standard deviation obtained using the normalized residuals distributions is $\sim 0.1\, \rm rad$ for the three codes. This uncertaininty does not effect our ability to perfrom APC test for current or near-future GW observations.
\begin{figure}[]
\includegraphics[width=0.98\columnwidth]{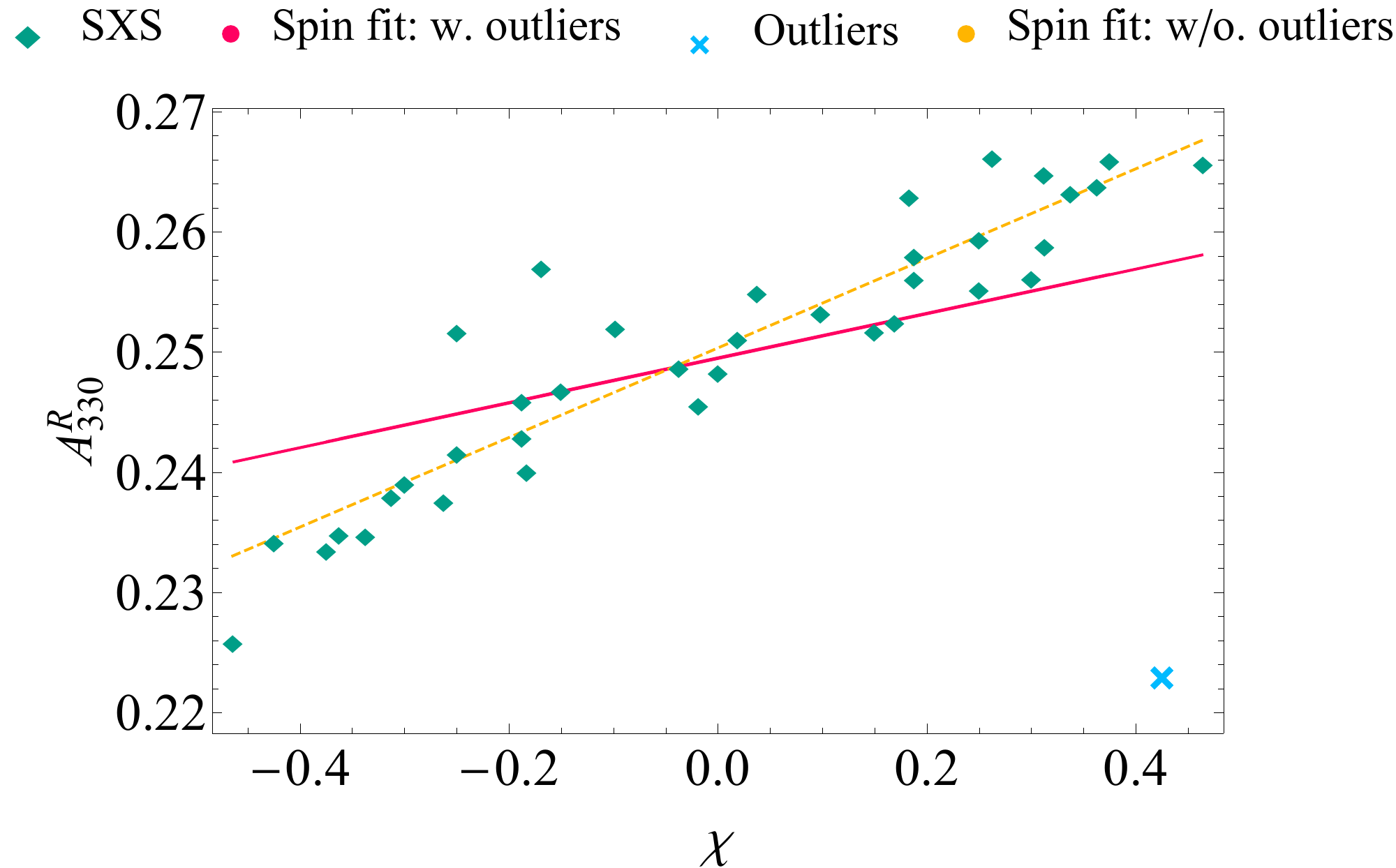}
    \caption{Amplitude ratio fits $A^R_{330}$ for $q=3$ in terms of the effective spin $\chi$. SXS data are denoted by green diamonds while the red dots denote the results obtained from fitting the data with the PN motivated ansatz~\eqref{eq:amp_ans} linear in $\chi$. One can easily identify an outlier point at $A^{R}_{330}=0.223$ and $\chi=0.850$ (blue cross). The solid orange curve provides the results of the linear fit after removing the outlier point. In particular, this corresponds to the SXS simulation SXS:0293.}
    \label{fig:outliers}
\end{figure}
\subsection{Identifying fit outliers in the data}
\label{sec:outliers}
We have observed that the BH spins have a minor impact on the values of $A^R_{lmn}$ and $\delta\phi_{lmn}$ compared to the effect of the mass ratio $q$. This allow us to use an ansatz linear in $\chi_{\rm pheno}$ for all the modes, except for  $\delta\phi_{210}$, in which the effects of $\chi_{\rm pheno}$ are found to be larger. While this property makes it easier to model $A^R_{lmn}$ and $\delta\phi_{lmn}$ in terms of the physical parameters, we find that presenced of outliers decrease the quality of the spin-dependent fits. The outliers seem to be dominated by numerical noise and we remove them from our calibration dataset by the following procedure.  We first look for highly significant outliers in both $A^R_{lmn}$ and $\delta\phi_{lmn}$ by testing the spin fit $\chi_{\rm pheno}$ at each mass ratio $q$. More specifically, we select the data at mass ratio $q=1,2,3,6,7,8$ to perform a bootstrap analysis over the $\chi_{\rm pheno}$ axis. This is achieved by computing the spinning fit at each $q$, that in general will contain $N$ points, for a data set of $N-1$ points. We iterate the fit for the $N$ points at each $q$, and we compute the value of the standard deviation $\sigma(q^i,\chi_{\rm pheno}^i)$. Then, we select the median value of $\bar{\sigma}$  and we discard all points beyond a conservative deviation of $4\bar{\sigma}$. In Fig.~\ref{fig:outliers} we show an example of this procedure applied to the amplitude ratio $A^R_{330}$. We see that the blue cross placed at the low-right corner is easily identified by this algorithm. However, it is important to keep a conservative criterion to avoid rejecting systematically false outliers.
\begin{figure}
\includegraphics[width=0.98\columnwidth]{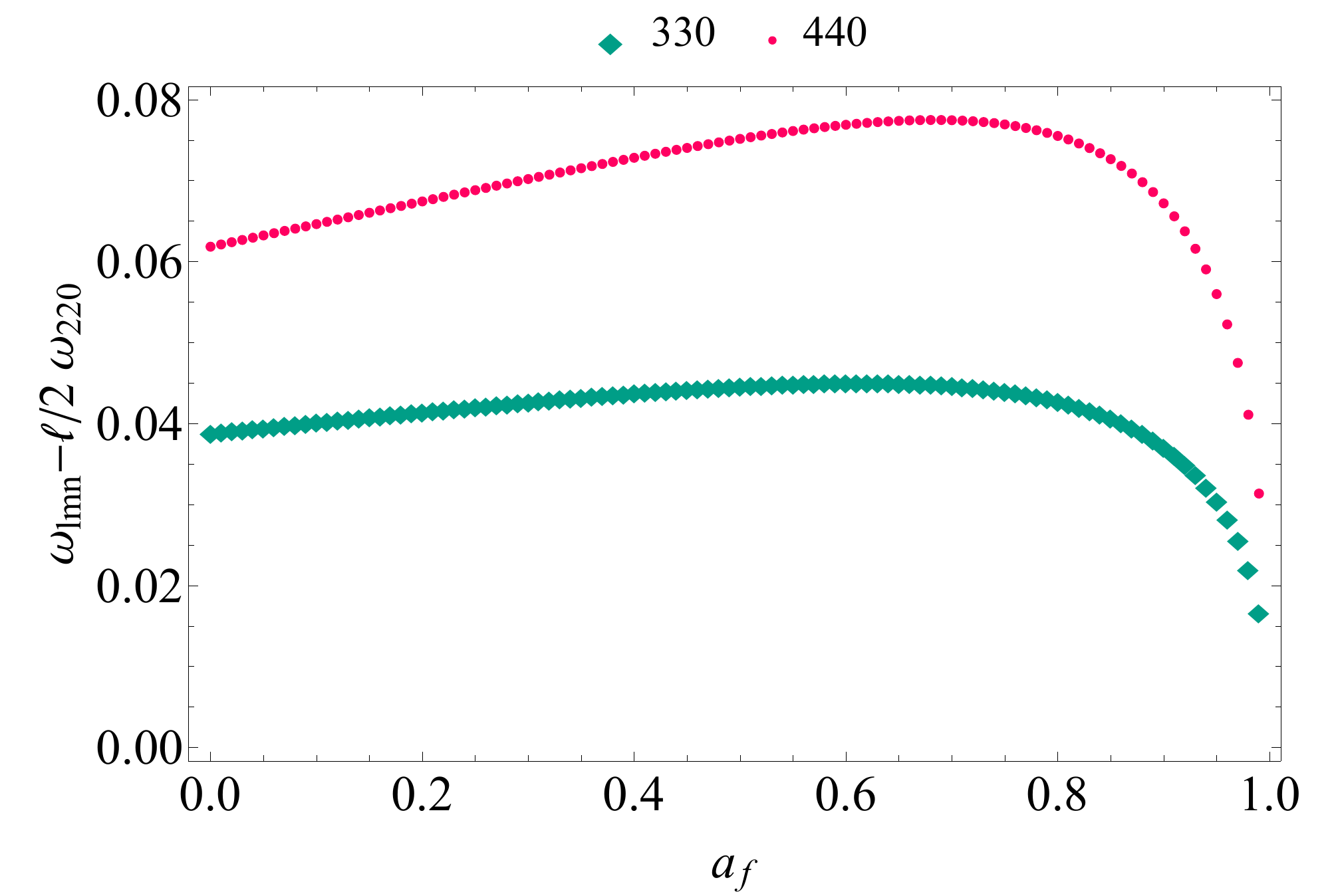}
    \caption{ Difference $\omega_{lmn}- l/2 \,\omega_{220}$ as a function of the remnant spin. The $\omega_{lmn}\sim l/2 \,\omega_{220}$ approximation is used in the calibration of $\delta\phi_{lmn}$ for the (3,3,0) and (4,4,0) modes. For the (3,3,0) mode and a conservative choice $\Delta t \sim 10M$ this adds an uncertainty of about $0.4\,\rm rad$. However, such approximation worsens for the ($4,4,0$) mode.}
    \label{fig:omega_disc}
\end{figure}
\subsection{On the $\Delta t$ dependence for the phase difference}
\label{ap:omegalmn_phase}
In order to refer the NR fits to the parameter estimation to a consistent initial time we need to shift each mode by $\Delta t=t^{\rm p}-t^{\rm p}_{220}$, where $t^{\rm p}$ is the global peak time of the signal and $t^{\rm p}_{220}$ is an arbitrary reference time taken to be the peak time of $|h_{220}|$. This time shift 
introduces a dephasing $\Delta \phi_{lmn} = \omega_{lmn} \Delta t$ and an extra term in the phase difference $\delta \phi_{lmn}(t^{\rm p})$ (see Eq.~\ref{dphaseshift})
\begin{equation}
    \left(\frac{m}{2}\omega_{220}-\omega_{lmn}\right)\Delta t\,.\label{extraterm}
\end{equation}
Fortunately, as can be shown analytically in the geodesics approximation valid in the eikonal limit $l=m\gg1$~\cite{ferrari:1984zz,Cardoso:2008bp}, for the fundamental ($n=0$) modes the following approximation holds: $\omega_{lm0}\sim l/2 \,\omega_{220}$. Thus, for the $l=m$ modes the extra terms in Eq.~\eqref{extraterm} is small. 
In Fig.~\ref{fig:omega_disc}, we show $\omega_{lmn}- l/2 \,\omega_{220}$ for the $(3,3,0)$ and $(4,4,0)$ modes as a function of the remnant spin, showing that the difference is in the range $\approx[0.04,0.08]$ for any spin.
This yields an ambiguity in $\delta \phi_{lmn}$ approximately of $(\omega_{330}\sim 3/2 \,\omega_{220})\Delta t \sim 0.04 \Delta t/M$ and $(\omega_{440}\sim 4/2 \,\omega_{220})\Delta t \sim 0.07 \Delta t/M$. Choosing a very conservative error estimate on $\Delta t$, namely $\Delta t =2(t^{\rm p}_{33}-t^{\rm p}_{22})= 10M$, we conclude that ignoring the dephasing introduced by $\Delta t$ will at most introduce an overall uncertainty $\delta\phi_{330}\sim 0.4\, \rm rad $ and $\delta\phi_{330}\sim 0.8\, \rm rad $. Note that, since $(t^{\rm p}_{33}-t^{\rm p}_{22})\sim 5M$ for all the NR simulations explored~\cite{forteza2020,Estelles:2020twz}, in practice we expect the uncertainty on $\delta\phi_{330}$ to be typically half of this conservative estimate.
\subsubsection{On the effect of the reference time for the parameter estimation}
\begin{figure}[]
    \includegraphics[width=\columnwidth]{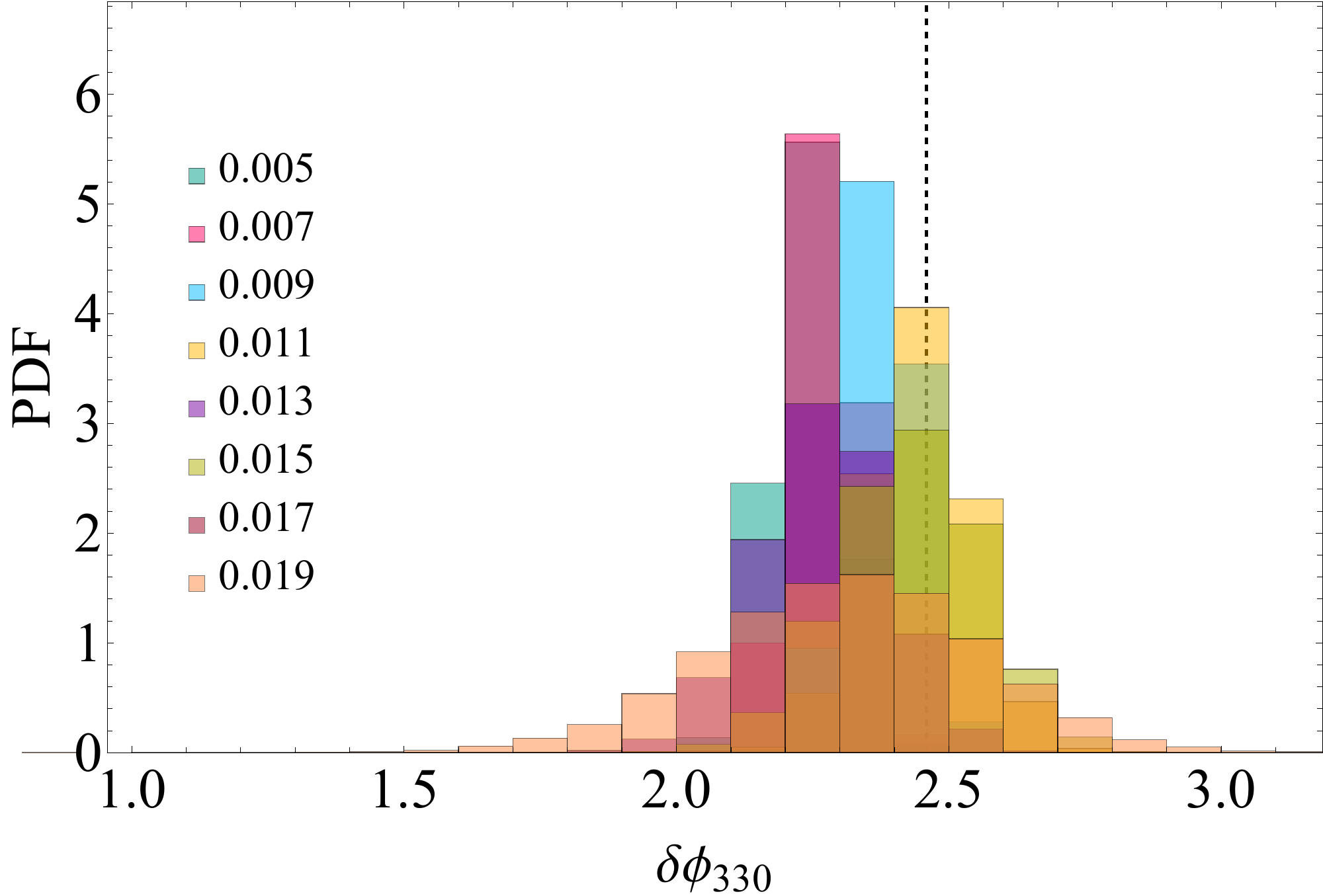}
    \caption{Posteriors distributions for $\delta\phi_{330}$ obtained from our injected signal SXS:0258 and for set of starting times $t_r\in \left[0.005,0.019\right] \rm ms$. The black dashed line stands for the fitting value shifted back $\Delta t (3/2 \omega_{220}-\omega_{330})$ to account for the $\Delta t=t^p-t^p_{220}$ difference. For this simulation and for an inclination $\iota=2.4$, we obtain $\Delta t=0.013 \rm ms$.}
    \label{fig:starting_time}
\end{figure}

We require that the reference time to start the parameter estimation on our GW190521 injections satisfies $t_r\geq t^p_{33}$. This  fixes a time at which the ($3,3,0$) mode is excited, which may source some systematic errors on $\delta\phi_{lmn}$.  For GW190521 with an inclination of $\iota=2.4$, we obtain $t_r\geq 0.005\, \rm ms$. In Fig.~\ref{fig:starting_time} we show the posterior distributions obtained for $\delta\phi_{330}$ for a set of reference times $\left[0.005,0.019\right] \rm ms$. The dashed black line corresponds to the fit value, shifted to $t^p_{22}$. As expected, the posterior distributions are consistent to each other for all the times $t_r$ considered here. 
\subsection{The effect of the eccentricity}\label{sec:eccentricity}
\begin{figure*}[]
    \includegraphics[width=0.48\textwidth]{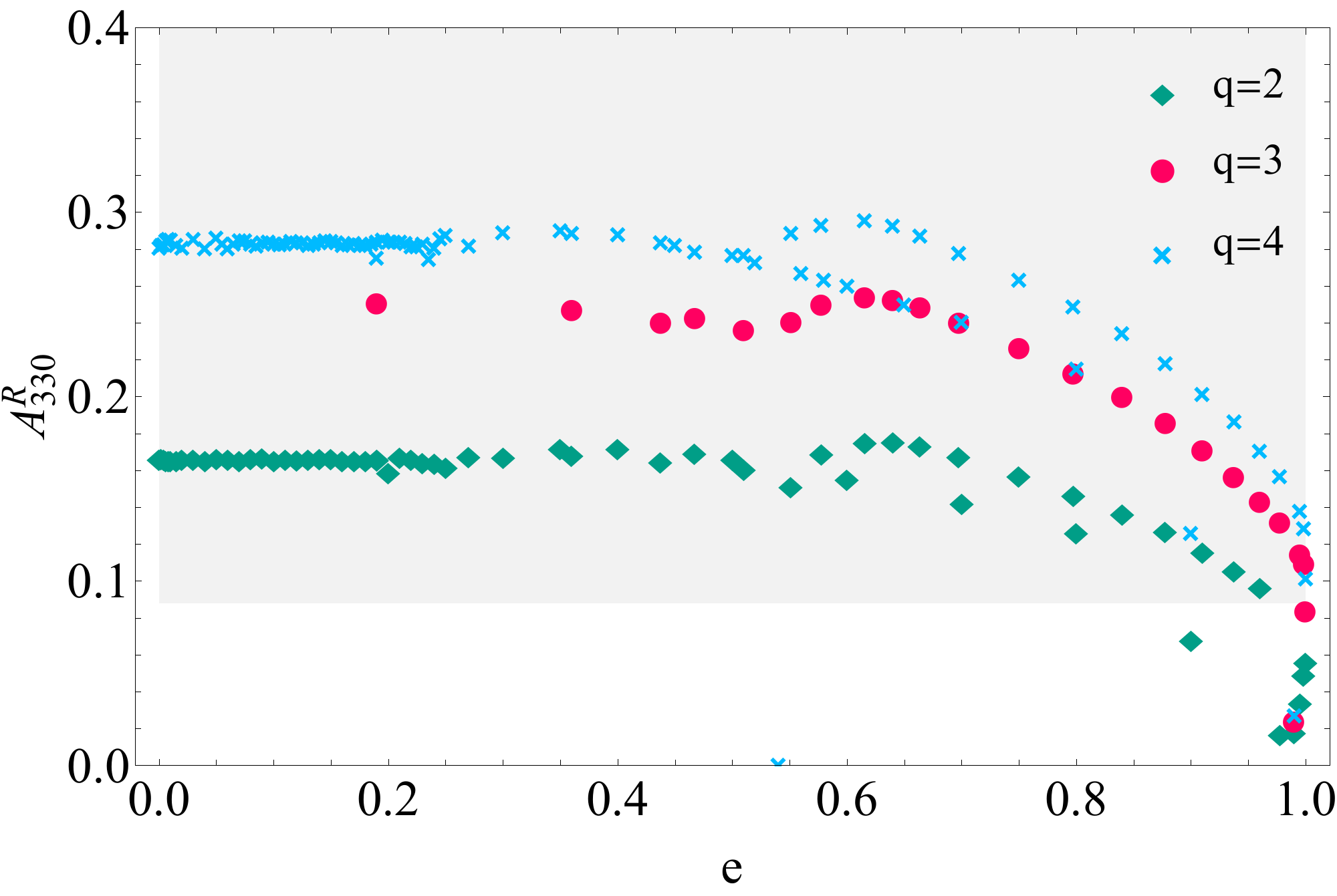}
    \includegraphics[width=0.48\textwidth]{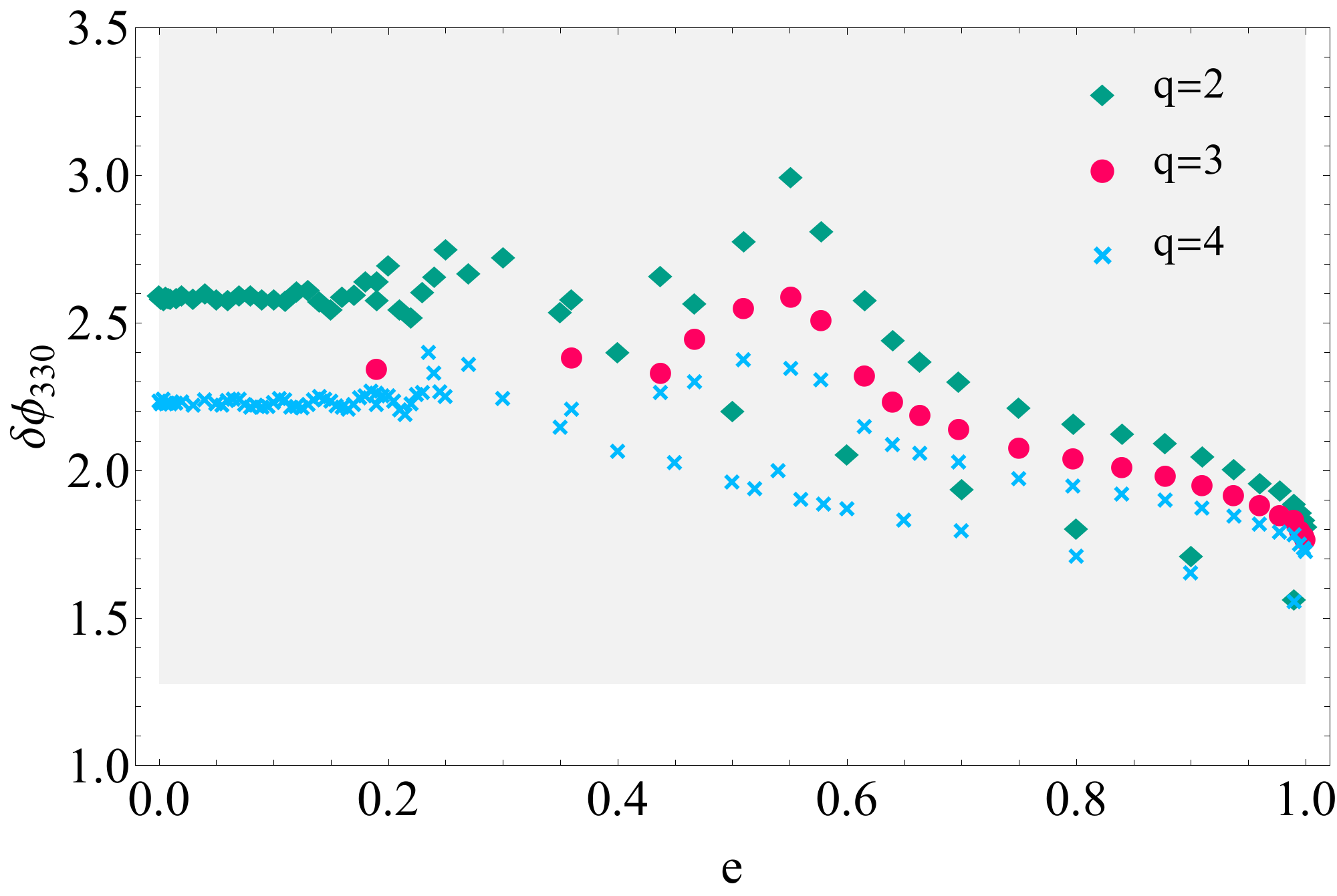}
    \caption{The impact of the eccentricity $e$ on the estimates of the amplitude ratio $A^R_{330}$ (left) and phase difference $\delta\phi_{330}$ (right) for a set of mass ratios $q=2,3,4$ and $\chi_{\rm pheno}=0$ using the RIT catalog~\cite{ritcatalog}. {The shaded area represents the $95\%$ credible intervals obtained from the posteriors shown in Fig.~\ref{fig:aratiophase}. Notice that, from the posterior distribution on $A^R_{330}$ and the NR fits of eccentric data, we can constrain the eccentricity $e\lesssim 0.9$ at the  $95\%$ confidence level. Moreover,} the effects of the eccentricity {will} become relevant only when $e\gtrsim 0.3$ for the $\delta\phi_{330}$ and when  $e\gtrsim 0.6$ for $A^R_{330}$ {and for GW events louder than GW190521}.}
    \label{fig:eccentricity}
\end{figure*}

The coalescence of eccentric binaries can in principle modify the initial perturbation conditions setup for the ringdown phase, and hence the QNM amplitudes and phases. Thus, the amplitude ratio $A^R_{lmn}$ and the phase difference $\delta\phi_{lmn}$ also depends on the eccentricity $e$.
In Fig.~\ref{fig:eccentricity} we examine the impact of the eccentricity on the values of $A^R_{330}$ and  $\delta\phi_{330}$  for a set of mass ratios $q=2,3,4$ -- where both the standard BH spectroscopy test and our APC test are promising~\cite{Forteza:2021wfq,Capano:2021etf}. To obtain these estimates, we use the data from the  RIT catalog~\cite{ritcatalog}. Notice that the values on $\delta\phi_{330}$ are  significantly modified only at relatively large values of the eccentricity with $e\gtrsim 0.3$, while this value raises up to $e\gtrsim 0.6$ for $A^R_{330}$. These values are still above the upper limit  threshold for $e~\sim 0.1$ obtained from the search of eccentric BBHs during the first and second LIGO observation runs~\cite{LIGOScientific:2019dag}. Thus eccentricities are not relevant for the most of the events observed by current ground based GW observatories for perfroming APC test. {Finally, the shaded area provides the $95\%$ credible intervals obtained from the posteriors shown in Fig.~\ref{fig:aratiophase}. Using these loose constraints, we obtain bounds on the eccentricity of GW190521 as $e\lesssim 0.9$ at the  $95\%$ confidence level.}

\section{Consistency between mode-excitations and BBH mass ratio in GW190521}\label{sec:IMR}
In this work we have proposed a new test  of GR  called the APC test that we demonstrate on GW190521 using the ringdown alone (and possibly prior knowledge of the binary extrinsic parameters~\cite{Baibhav:2020tma}). In this section, we highlight another possible null test of GR that makes use of the amplitude ratio and its relation to the BBH mass ratio, and therefore requires the entire inspiral-merger-ringdown~(IMR) signal. The basic concept here is that one could estimate the mass ratio $q$ from the ringdown by inverting the $A^{R}_{lmn} = A^{R}_{lmn} (q)$ relation and then check whether the inferred value is consistent with $q$ measured independently from the full IMR signal\footnote{While in principle an IMR consistency test can be done by directly checking the consistency of the fundamental mode amplitude $A_{220}$ as a function of the binary parameters, this quantity depends on several (both intrinsic and extrinsic) parameters so its constraining power is limited, e.g., by correlations. However, the amplitude ratio $A_{lmn}^R$ depends mainly on the binary mass ratio and spins.}. This is complementary to the standard IMR consistency tests performed by the LVK Collaboration~\cite{LIGOScientific:2021sio} and is based on an idea similar to that used to design the merger-ringdown test for the BBH population presented in Ref.~\cite{Bhagwat:2021kwv}.  

Our fits provide empirically $2$-to-$2$ maps $(A^R_{lmn},\delta\phi_{lmn})\to (q,\chi_{\rm pheno})$ which can be inverted to obtain $q$. Note that the sensitivity of this test toward the measurement of the spin is limited since $A^{R}_{330}$ and $A^{R}_{440}$ have a rather mild dependence on the spins (see Fig.~\ref{fig:aratio_fits_q}). 
However, even when one neglects the spin dependence, one can map $A^{R}$ to $q$ to a good approximation. For instance, for $q=2$ we get $A^R_{330}=0.14,0.19$ for $\chi_{1,2}=\pm 0.85$; the spin dependence is subleading. In principle, this kind of test could also be designed with $\delta \phi_{lmn}$. However, in practice the dependence of $\delta \phi_{lmn}$ on the BBH parameters is weak, and one expects a much larger errors on the inferred BBH parameters making the test less constraining.

\begin{figure}
\includegraphics[width=0.98\columnwidth]{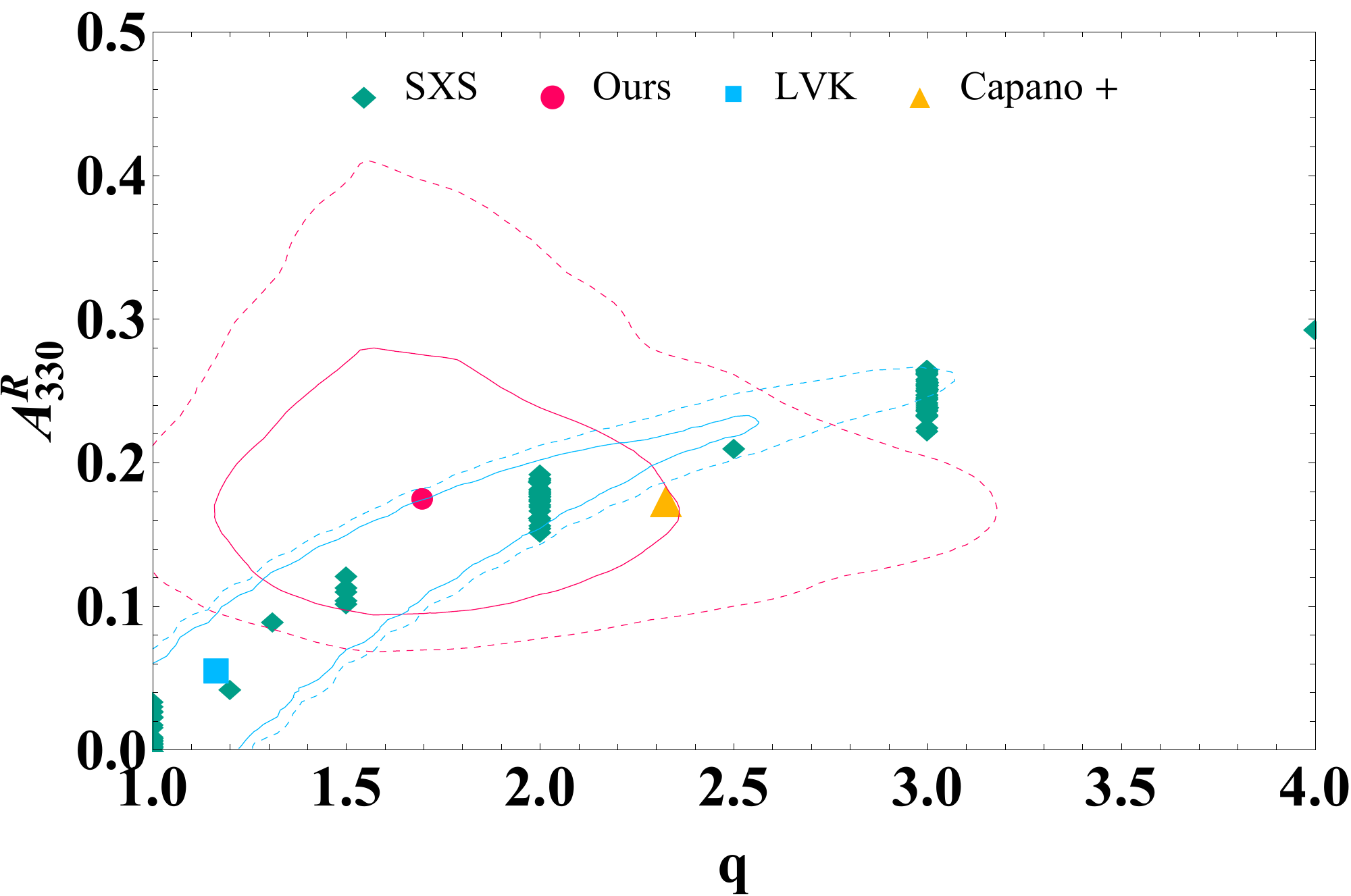}
    \caption{
    Amplitude ratio $A^R_{330}$ estimated from the set of NR waveforms used in this work and restricted to $q<3$. The mild dependence on the spin parameters is reflected by the relatively small spread of the points at fixed $q$. We add on top the different estimates of the $A^R_{330}-q$ relation obtained from different mass ratio estimates: Capano+ (orange triangle), Ours (red dot) and LVK (blue square). The values of $A^R_{330}$ and $q$ predicted from NR are consistent at the $2\sigma$ level with the distributions obtained on $q$ and $A^R_{lmn}$ obtained from~\cite{LIGOScientific:2021djp} and~\cite{Capano:2021etf}.}
    \label{fig:aratio_q}
\end{figure}

We scrutinize GW190521 for consistency between mode-excitation and BBH mass ratio. Owing to its large total mass and short inspiral signal in the LIGO-Virgo band, the inspiral parameter estimation of GW190521 is particularly sensitive to model systematics and there is some tension among the binary parameters (including the mass ratio) inferred with different waveforms~\cite{Nitz:2020mga,Kastha:2021chr,LIGOScientific:2021djp}. Thus, one cannot perform a reliable IMR-like test on this signal. In Fig.~\ref{fig:aratio_q} we compare the different estimates for $q$ using parameter estimation posteriors provided in these IMR studies with the expected value of $q$ inferred from ringdown. For the ringdown estimate of $q$, we compare the estimate of $A^R_{330}$ obtained in~\cite{Capano:2021etf} along with our fit results. Specifically, we show
\begin{enumerate}[label=(\roman*)]

    \item {the parameter-estimation results on $A^R_{lmn}$ from~\cite{Capano:2021etf} (red) and translated to $q$ by inverting our $A^R_{330}$ mode fit. The red dot (Ours) provides its best likelihood value obtained from the marginalized distribution ${q-A^R_{330}}$.}
    \item {the parameter-estimation results on $q$ and $\chi_{1,2}$ obtained by the LVK collaboration~\cite{LIGOScientific:2021djp} (blue) and translated to $A^R_{lmn}$ by using our ($3,3,0$) mode fit. The blue-square (LVK) provides its best likelihood value obtained from the marginalised distribution ${q-A^R_{330}}$.} Since the LVK does not provide an independent distribution on $A^R_{330}$, we get the elongated blue contours using our fit. 
\end{enumerate}
We observe that Ours, Capano+, and LVK are all consistent with the parameter-estimation posterior distribution at the $1\sigma$ and $2\sigma$ credible level. Note that the $q$ distribution from Capano+ is obtained from an independent set of fits~\cite{borhanian:2019kxt}. We also obtained a rather flat distribution on the phenomenological spin parameter with  $\chi \in\left[0,1\right]$, which is expected given the mild dependence of both $A^R_{330}(q,\chi_{\rm pheno})$ and $\delta\phi_{330}(q,\chi_{\rm pheno})$ on $\chi$. However, this paradigm may change shortly with louder detections such as those expected from third-generation detectors~\cite{Maggiore:2019uih,Kalogera:2021bya} and LISA~\cite{Bhagwat:2021kwv}. 
\end{document}